\documentclass[journal]{IEEEtran}

\usepackage{amssymb,epic,eepic}

\usepackage[dvips]{color}
\usepackage{graphicx}

\usepackage{pst-plot, pstricks}

\begin{document}
\title{Strongly Universal Quantum Turing Machines and Invariance of
Kolmogorov Complexity}

\newcommand{\hr}{{\cal H}}
\newcommand{\cnq}{{\hr_n^\mathbb{Q}}}
\newcommand{\cn}{{\hr_n}}
\newcommand{\C}{{\mathbb C}}
\newcommand{\R}{{\mathbb R}}
\newcommand{\N}{{\mathbb N}}
\newcommand{\idn}{\mathbf{1}}
\newcommand{\Z}{{\mathbb Z}}
\newcommand{\x}{{\mathbf x}}
\newcommand{\eps}{{\varepsilon}}
\newcommand{\om}{\omega}
\newcommand{\kk}{{\mathbf k}}
\newcommand{\n}{{\mathbf n}}
\newcommand{\y}{{\mathbf y}}
\newcommand{\s}{{\{0,1\}^*}}
\newtheorem{theorem}{Theorem}[section]
\newtheorem{lemma}[theorem]{Lemma}
\newtheorem{corollary}[theorem]{Corollary}
\newtheorem{definition}[theorem]{Definition}
\newtheorem{proposition}[theorem]{Proposition}
\newtheorem{conjecture}[theorem]{Conjecture}
\newtheorem{example}[theorem]{Example}
\newcommand{\nix}{{\rule{0pt}{2pt}}}
\newcommand{\qedd}{{\nix\nolinebreak\hfill\hfill\nolinebreak$\Box$}}
\newcommand{\qed}{{\qedd\par\medskip\noindent}}
\newcommand{\lineclear}{{\rule{0pt}{0pt}\nopagebreak\par\nopagebreak\noindent}}

\author{Markus M\"uller%
\thanks{M. M\"uller is with the Institute of Mathematics, Technical University of Berlin (e-mail:
mueller@math.tu-berlin.de).}}%
%
\markboth{Strongly Universal Quantum Turing Machines and Invariance
of Kolmogorov Complexity (August 11, 2007)}{}

\maketitle

\begin{abstract}
We show that there exists a universal quantum Turing machine (UQTM) that can simulate
every other QTM {\em until the other QTM has halted} and then halt itself with probability one.
This extends work by Bernstein and Vazirani who
have shown that there is a UQTM that can simulate every other QTM
for an arbitrary, but preassigned number of time steps.

As a corollary to this result, we give a rigorous proof that quantum Kolmogorov complexity as
defined by Berthiaume et al. is invariant, i.e. depends on the choice of the
UQTM only up to an additive constant.

Our proof is based on a new mathematical framework for QTMs, including
a thorough analysis of their halting behaviour. We introduce the notion of
mutually orthogonal halting spaces and show 
that the information encoded in an input
qubit string can always be effectively decomposed
into a classical and a quantum part.
\end{abstract}


%
\IEEEpeerreviewmaketitle

\section{Introduction}
\PARstart{O}{ne} of the fundamental breakthroughs of computer science was the insight that there
is a single computing device, the universal Turing machine (TM), that can simulate every
other possible computing machine. This notion of universality laid the foundation
of modern computer technology. Moreover, it provided the opportunity to study
general properties of computation valid for every possible computing device at once,
as in computational complexity and algorithmic information theory respectively.

Due to the development of quantum information theory in recent years, much work has been done
to generalize the concept of universal computation to the quantum realm.
In 1985, Deutsch~\cite{Deutsch} proposed
the first model of a quantum Turing machine (QTM), elaborating on an
even earlier idea by Feynman~\cite{Feynman}. Bernstein and Vazirani~\cite{BernsteinVazirani}
worked out the theory in more detail and proved that there exists a QTM that is universal in the
sense that it efficiently simulates every other possible QTM. This remarkable result provides
the foundation to study quantum computational complexity, especially the complexity class BQP.

In this paper, we shall show that there exists a QTM that is universal in the sense of
program lengths. This is a different notion of universality, which is needed to study quantum
algorithmic information theory.
The basic difference is that the ``strongly universal'' QTM constructed in this paper
does not need to know the number of time steps of the computation in advance, which is difficult
to achieve in the quantum case.

For a compact presentation of the results by Bernstein and Vazirani, see the book by Gruska~\cite{Gr}.
Additional relevant literature includes
Ozawa and Nishimura~\cite{Ozawa}, who
gave necessary and sufficient conditions that a QTM's transition function results in
unitary time evolution. Benioff~\cite{Benioff} has worked out a slightly
different definition which is based on a local Hamiltonian instead of a local transition amplitude.

\subsection{Quantum Turing Machines and their Halting Conditions}
\label{SubsecIntroQTM}
Our discussion will rely on the definition by Bernstein and Vazirani. We describe their
model in detail in Subsection~\ref{SubsecQTMs}.
Similarly to a classical TM\footnote{We use the terms ``Turing machine'' (TM) and ``computer''
synonymously for ``partial recursive function from $\{0,1\}^*$ to $\{0,1\}^*$'', where
$\{0,1\}^*=\{\lambda,0,1,00,\ldots\}$ denotes the finite binary strings.
}, a QTM consists of an infinite tape, a control,
and a single tape head that moves along the tape cells. The QTM as a whole
evolves unitarily in discrete time steps. The (global) unitary time evolution $U$ is completely
determined by a local transition amplitude $\delta$ which only affects the single tape cell where
the head is pointing to.

There has been a vivid discussion in the literature on the question when we
can consider a QTM as having {\em halted} on some input and how this is compatible
with unitary time evolution, see e.g. \cite{Myers, LindenPopescu, OzawaLocalTransition, Shi, Miyadera}.
We will not get too deep into this discussion, but rather
analyze in detail the simple definition for halting by Bernstein and Vazirani \cite{BernsteinVazirani},
which we also use in this paper. We argue below that
this definition is useful and natural, at least for the purpose to study quantum Kolmogorov complexity.

Suppose a QTM $M$ runs on some quantum input $|\psi\rangle$ of $n$ qubits for $t$ time steps.
The control $\mathbf C$ of $M$ will then be in some state (obtained by partial trace over the
all the other parts of the QTM) which we denote $M_{\mathbf C}^t(|\psi\rangle)$. In general,
this is some mixed state on the finite-dimensional Hilbert space $\hr_{\mathbf C}$
that describes the control. By definition of a QTM (see Subsection~\ref{SubsecQTMs}), there is a specified
{\em final state} $|q_f\rangle\in\hr_{\mathbf C}$.
According to \cite{BernsteinVazirani},
we say that the QTM $M$ halts at time $T$ on input $|\psi\rangle$ if
\[
   \langle q_f|M_{\mathbf C}^T(|\psi\rangle)|q_f\rangle=1
   \mbox{ and }
   \langle q_f|M_{\mathbf C}^t(|\psi\rangle)|q_f\rangle=0\quad\forall t<T.
\]
We can rephrase this definition as $M_{\mathbf C}^T(|\psi\rangle)=|q_f\rangle\langle q_f|$, i.e.
the control is {\em exactly} in the final state at time $T$, and ${\rm supp}\left(M_{\mathbf C}^t(|\psi\rangle)
\right)\perp  |q_f\rangle$, i.e. the control state is exactly orthogonal to the halting state
at any time $t<T$ before the halting time.

In general, the overlap of $M_{\mathbf{C}}^t(|\psi\rangle)$ with the final state $|q_f\rangle$ will be
some arbitrary number {\em between} zero and one. Hence, for most input qubit strings $|\psi\rangle$,
there will be no time $T\in\N$ such that the aforementioned halting conditions are satisfied. We call
those qubit strings {\em non-halting}, and otherwise
{\em $T$-halting}, where $T\in\N$ is the corresponding halting time.

In Subsection~\ref{SubsecHaltingSubspaces}, we analyze the resulting geometric structure
of the halting input qubit strings. We show that inputs $|\psi\rangle$ with
some fixed length $n$ that make the QTM $M$ halt after $t$ steps form a linear
subspace $\hr_M^{(n)}(t)$. Moreover,
inputs with different halting times are mutually orthogonal, i.e.
$\hr_M^{(n)}(t)\perp \hr_M^{(n)}(t')$ if $t\neq t'$. According to the halting conditions given
above, this is almost obvious: Superpositions of $t$-halting inputs are again $t$-halting,
and inputs with different halting times can be perfectly distinguished, just
by observing their halting time.

\begin{figure}[!hbt]
\begin{center}
\includegraphics[angle=0, width=5cm]{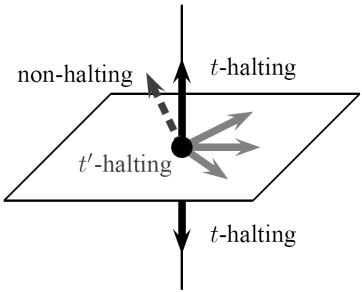}
\caption{mutually orthogonal halting spaces.}
\label{AbbHaltingSpaces}
\end{center}
\end{figure}

In Figure~\ref{AbbHaltingSpaces}, a geometrical
picture of the halting space structure is shown: The whole space $\R^3$ represents the space
of inputs of some fixed length $n$, while the plane and
the straight line represent two different halting spaces $\hr_M^{(n)}(t')$ and $\hr_M^{(n)}(t)$.
Every vector within these subspaces is perfectly halting, while every vector ``in between''
is non-halting and not considered a useful input for the QTM $M$.


At first, it seems that the halting conditions given above are far too restrictive. Don't we
loose a lot by dismissing every input which does not satisfy those conditions perfectly,
but, say, only approximately up to some small $\eps$? To see that it is not that bad, note that
\begin{itemize}
\item most (if not all) of the well-known quantum algorithms, like the quantum Fourier transform
or Shor's algorithm, have classically controlled halting. That is, the halting time is known
in advance, and can be controlled by a classical subprogram.
\item we show elsewehere~\cite{MeineDiss} (cf. Theorem~\ref{Conjecture})
that every input that is {\em almost}
halting can be modified by adding at most a constant number of qubits to halt {\em perfectly},
i.e. to satisfy the aforementioned halting conditions. This can be interpreted as some kind
of ``stability result'', showing that the halting conditions are not ``unphysical'', but
have some kind of built-in error tolerance that was not expected from the beginning.
\end{itemize}

Moreover, this definition of halting is very useful. Given two QTMs $M_1$ and $M_2$, it enables
us to construct a QTM $M$ which carries out the computations of $M_1$, {\em followed by the
computations of $M_2$}, just by redirecting the final state $|q_f\rangle$ of $M_1$ to
the starting state $|q_0\rangle$ of $M_2$ (see \cite[Dovetailing Lemma 4.2.6]{BernsteinVazirani}).
In addition, it follows from this definition that QTMs are {\em quantum operations},
which is a very useful and plausible property.

Even more important, at each single time
step, an outside observer can make a measurement of the control state, described by
the operators $|q_f\rangle\langle q_f|$ and $\idn-|q_f\rangle\langle q_f|$ (thus observing
the halting time), without
spoiling the computation, as long as the input $|\psi\rangle$ is halting. As soon as
halting is detected, the observer can extract the output quantum state from the output track (tape)
and use it for further quantum information processing.
This is true even if the halting time
is very large, which typically happens in the study of Kolmogorov complexity.
Consequently, our definition of halting has the useful property that if an outside
observer is given some unknown quantum state $|\psi\rangle$ which is halting, then the observer
can find out with certainty by measurement.

Finally, if we instead introduced some probabilistic notion of halting (say, we demanded
that we observe halting of the QTM $M$ at some time $t$ with some large probability $p<1$),
then it would not be so clear how to define quantum Kolmogorov complexity correctly. Namely
if the halting probability is much less than one, it seems necessary to introduce some kind
of ``penalty term'' into the definition of quantum Kolmogorov complexity: there should
be some trade-off between program length and halting accuracy, and it is not so clear
what the correct trade-off should be. For example, what is the complexity of a qubit string
that has a program of length 100 which halts with probability $0.8$, and another program
of length 120 which halts with probability $0.9$? The definition of halting that we use
in this paper avoids such questions.

\subsection{Different Notions of Universality for QTMs}
\label{SubsecUniversalities}
Bernstein and Vazirani~\cite{BernsteinVazirani} have shown that there exists a universal QTM
(UQTM) $\mathcal{U}$. It is important to understand what exactly they mean by ``universal''.
According to \cite[Thm. 7.0.2]{BernsteinVazirani}, this UQTM $\mathcal U$ has the property
that for every QTM $M$
there is some classical bit string $s_M\in\{0,1\}^*$ (containing a description of
the QTM $M$) such that
\begin{equation}
   \left\|\mathcal{U}(s_M,T,\delta,|\psi\rangle)-\mathcal{R}\left(M_{\mathbf O}^T(|\psi\rangle)\right)\right\|_{\rm Tr}<\delta
   \label{EqWeakUniversality}
\end{equation}
for every input $|\psi\rangle$, accuracy $\delta>0$ and number of time steps $T\in\N$.
Here, $\|\cdot\|_{\rm Tr}$ is the trace distance, and $\mathcal{R}\left(M_{\mathbf O}^T(|\psi\rangle)\right)$ is the
content of the output tape $\mathbf O$ of $M$ after $T$ steps of computation (the notation will be defined
exactly in Subsection~\ref{SubsecQTMs}).

This means that the UQTM $\mathcal U$ simulates every other QTM $M$ within any desired accuracy
and outputs an approximation of the output track content of $M$ and halts,
as long as the number of time steps $T$ is given as input in advance.

Since the purpose of
Bernstein and Vazirani's work was to study the computational complexity of QTMs,
it was a reasonable assumption that the halting time $T$ is known in advance (and not too large)
and can be specified as additional input. The most important point for them was not to have short inputs, but
to prove that the simulation of $M$ by $\mathcal U$ is efficient, i.e. has only polynomial slowdown.

The situation is different if one is interested in studying quantum algorithmic information theory
instead. It will be explained in Subsection~\ref{SubsecInvariance}
below that the universality notion (\ref{EqWeakUniversality}) is not enough for proving
the important invariance property of quantum Kolmogorov complexity, which says that quantum
Kolmogorov complexity depends on the choice of the universal QTM only up to an additive constant.

To prove the invariance property, one needs a generalization 
of (\ref{EqWeakUniversality}), where the requirement to have the running time $T$ as
additional input is dropped. We show below in Section~\ref{SecConstruction}
that there exists a UQTM $\mathfrak U$ that satisfies such a generalized
universality property, i.e. that simulates every other QTM until that other
QTM has halted, without knowing that halting time in advance, and then halts itself.

Why is that so difficult to prove? At first, it seems that one can just program
the UQTM $\mathcal U$  mentioned in (\ref{EqWeakUniversality})
to simulate the other QTM $M$ for $T=1,2,3,\ldots$ time steps, and,
after every time step, to check if the simulation of $M$ has halted or not. If it has halted,
then $\mathcal U$ halts itself and prints out the output of $M$, otherwise it continues.

This approach works for classical TMs, but for QTMs, there is one problem: in general,
the UQTM $\mathcal U$ can simulate $M$ only approximately. The reason is the same as for the circuit
model, i.e. the set of basic unitary transformations that $\mathcal U$ can apply on its
tape may be algebraically independent from that of $M$, making a perfect simulation in principle
impossible. But if the simulation is only approximate, then the control state
of $M$ will also be simulated only approximately, which will force
$\mathcal U$ to halt only approximately. Thus, the restrictive halting conditions given
above in Equation~(\ref{EqHalting}) will inevitably be violated,
and the computation of $\mathcal U$ will be treated as invalid and be dismissed by definition.

This is a severe problem that cannot be circumvented easily. Many ideas for simple solutions
must fail, for example the idea to let $\mathcal U$ compute an upper bound on the halting time
$T$ of all inputs for $M$ of some length $n$ and just to proceed for $T$ time steps: upper bounds on
halting times are not computable. Another idea is that the computation of $\mathcal U$ should
somehow consist of a classical part that controls the computation and a quantum part that
does the unitary transformations on the data. But this idea is difficult to formalize.
Even for classical TMs, there is no general way to split the computation into ``program'' and
``data'' except for special cases, and for QTMs, by definition, global unitary time evolution
can entangle every part of a QTM with every other part.

Our proof idea rests instead on the observation that every {\em input} for a QTM which is
halting can be decomposed
into a classical and a quantum part, which is related to the mutual orthogonality of the
halting spaces. See Subsection~\ref{SubsecProofIdeas} for details.

\subsection{Q-Kolmogorov Complexity and its Supposed Invariance}
\label{SubsecInvariance}
The classical Kolmogorov complexity $C_U(s)$ of a finite bit string
$s\in\{0,1\}^*$ is defined as the minimal length of any computer program $p$ that,
given as input into a TM $M$, outputs the string and makes $M$ halt:
\[
   C_M(s):=\min\left\{\ell(p)\,\,|\,\,M(p)=s\right\}.
\]
For this quantity, running times are not important; all that matters is the input length.
There is a crucial result that is the basis for the whole theory of Kolmogorov complexity
(see \cite{Vitanyibook}). Basically, it states that the choice of the computer $M$
is not important as long as $M$ is universal; choosing a different
universal computer will alter the complexity only up to some
additive constant. More specifically, there exists a universal computer $U$ such that
for every computer $M$ there is a constant $c_M\in\N$ such that
\begin{equation}
   C_U(s)\leq C_M(s)+c_M\qquad\mbox{for every }s\in\{0,1\}^*.
   \label{eqInvClass}
\end{equation}
This so-called ``invariance property'' follows easily from the existence of a
universal computer $U$ in the following sense: There exists a computer $U$ such
that for every computer $M$ and every input $s\in\{0,1\}^*$ there is an input $\tilde s\in\{0,1\}^*$
such that $U(\tilde s)=M(s)$ and $\ell(\tilde s)\leq \ell(s)+c_M$, where $c_M\in\N$ is
a constant depending only on $M$. In short, there is a computer $U$ that produces every
output that is produced by any other computer,
while the length of the corresponding input blows up only by a constant summand.
One can think of the bit string $\tilde s$ as consisting of the original bit string $s$
and of a description of the computer $M$ (of length $c_M$).

The quantum generalization of Kolmogorov complexity that we consider in this paper has
been first defined by Berthiaume, van Dam and Laplante \cite{Berthiaume}. Basically, they define the quantum
Kolmogorov complexity $QC$ of a string of qubits $|\psi\rangle$ as the length of the shortest string of qubits
that, when given as input to a QTM $M$, makes $M$ output $|\psi\rangle$ and halt. (We give
a formal definition of a ``qubit string'' in Subsection~\ref{SubsecQubitStrings} and
of quantum Kolmogorov complexity $QC$ in Subsection~\ref{SubsecQComplexity}).

In \cite{Berthiaume}, it is claimed that quantum
Kolmogorov complexity $QC$ is invariant up to an additive constant similar
to (\ref{eqInvClass}). It is stated there that
the existence of a universal QTM $\mathcal U$ in the sense
of Bernstein and Vazirani (see Equation~(\ref{EqWeakUniversality}))
makes it possible to mimic the classical proof and to conclude that the UQTM $\mathcal U$
outputs all that every other QTM outputs, implying invariance of quantum Kolmogorov complexity.

But this conclusion cannot be drawn so easily, because (\ref{EqWeakUniversality}) demands that the halting time $T$ is specified
as additional input, which can enlarge the input length dramatically, if $T$ is very large
(which typically happens in the study of Kolmogorov complexity).

As explained
above in Subsection~\ref{SubsecUniversalities}, it is not so easy to get rid of the halting time.
The main reason is that the UQTM $\mathcal U$ can simulate other QTMs only approximately.
Thus, it will also simulate the control state and the signaling of halting only approximately,
and cannot just ``halt whenever the simulation has halted'', because then, it will
violate the restrictive halting conditions given in Equation~(\ref{EqHalting}).
As we have chosen this definition of halting for good reasons (cf. the discussion at the beginning
of Subsection~\ref{SubsecIntroQTM} above), we do not want to drop it.

Instead of (\ref{EqWeakUniversality}), a stronger notion of universality is needed,
namely a ``strongly universal'' QTM $\mathfrak U$ that, as explained above in
Subsection~\ref{SubsecUniversalities}, simulates every other QTM $M$ {\em until the other QTM
has halted} and then halts itself with probability one, as required by the halting
conditions given in Subsection~\ref{SubsecIntroQTM}. Then, the classical proof outlined above
can be carried over to the quantum situation. In this paper, we prove that such a QTM $\mathfrak U$
really exists (Theorem~\ref{MainTheorem1}), and as a corollary, the invariance property
for quantum Kolmogorov complexity follows (Theorem~\ref{MainTheorem2}).

\subsection{Main Theorems}
One main result of this paper is the existence of a ``strongly universal'' QTM that simulates
every other QTM until the other QTM has halted and then halts itself. Note that
the halting state is attained by $\mathfrak U$ {\em exactly} (with probability one) in accordance
with the strict halting conditions stated in Equation~(\ref{EqHalting}).
The exact definition of ``qubit strings'' and the output $M(\sigma)$ of $M$ on input $\sigma$
is given below in Section~\ref{SecFramework}.
\begin{theorem}[Strongly Universal Q-Turing Machine]
\label{MainTheorem1}
There is a fixed-length quantum Turing machine $\mathfrak{U}$ such that
for every QTM $M$ and
every qubit string $\sigma$
for which $M(\sigma)$ is defined, there is a qubit string $\sigma_M$
such that
\[
   \left\| \mathfrak{U}\,(\sigma_M,\delta)-M(\sigma)\right\|_{\rm Tr} <\delta
\]
for every $\delta\in\mathbb{Q}^+$, where the length of $\sigma_M$ is bounded
by $\ell(\sigma_M)\leq\ell(\sigma)+c_M$, and $c_M\in\N$ is a constant depending only
on $M$.
\end{theorem}
Note that $\sigma_M$ does not depend on $\delta$.
We conclude from this theorem and a two-parameter generalization
given in Proposition~\ref{PropTwoParameter} that quantum Kolmogorov complexity
as defined in \cite{Berthiaume} is indeed invariant, i.e. depends on the choice
of the strongly universal QTM only up to some constant:

\begin{theorem}[Invariance of Q-Kolmogorov Complexity]
\label{MainTheorem2}
There is a fixed-length quantum Turing machine $\mathfrak{U}$ such
that for every QTM $M$ there is a constant $c_M\in\N$ such that
\[
   QC_\mathfrak{U}(\rho)\leq QC_M(\rho)+c_M\qquad
   \mbox{for every qubit string }\rho.
\]
Moreover, for every QTM $M$ and every $\delta,\Delta\in\mathbb{Q}^+$ with $\delta<\Delta$,
there is a constant $c_{M,\delta,\Delta}\in\N$ such that
\[
   QC^\Delta_{\mathfrak{U}}(\rho)\leq QC^\delta_M(\rho)+c_{M,\delta,\Delta}
   \qquad\mbox{for every qubit string }\rho.
\]
\end{theorem}
All the proofs are given in Section~\ref{SecConstruction}, while the ideas of the proofs
are outlined in the next subsection.

\subsection{Ideas of Proof}
\label{SubsecProofIdeas}
The proof of Theorem~\ref{MainTheorem1} relies on the observation about the mutual orthogonality
of the halting spaces, as explained in Subsection~\ref{SubsecIntroQTM}. Fix some QTM $M$, and
denote the set of vectors $|\psi\rangle\in\left(\C^2\right)
^{\otimes n}$ which cause $M$ to halt at time $t$ by $\hr_M^{(n)}(t)$. If $|\varphi\rangle\in
\left(\C^2\right)^{\otimes n}$ is any halting input for $M$, then we can decompose $|\varphi\rangle$
in some sense into a classical and a quantum part. Namely, the information contained in $|\varphi\rangle$
can be split into a
\begin{itemize}
\item classical part: The vector $|\varphi\rangle$ is an element of
{\em which} of the subspaces $\hr_M^{(n)}(t)$?
\item quantum part: Given the halting time $\tau$ of $|\varphi\rangle$,
then {\em where} in the corresponding subspace $\hr_M^{(n)}(\tau)$
is $|\varphi\rangle$ situated?
\end{itemize}
Our goal is to find a QTM $\mathfrak U$ and an encoding
$|\tilde\varphi\rangle\in\left(\C^2\right)^{\otimes (n+1)}$
of $|\varphi\rangle$
which is only one qubit longer and which makes the (cleverly programmed) QTM $\mathfrak U$ output a good
approximation of $M(|\varphi\rangle)$. First, we extract the quantum part out of $|\varphi\rangle$.
While $\dim \left(\C^2\right)^{\otimes n}=2^n$, the halting space $\hr_M^{(n)}(\tau)$ that
contains $|\varphi\rangle$ is only a subspace and might have much smaller dimension $d<2^n$.
This means that we need less than $n$ qubits to describe the state $|\varphi\rangle$; indeed,
$\lceil \log_2 d\rceil$ qubits are sufficient. In other words, there is some kind of ``standard
compression map'' $\mathcal C$ that maps every vector $|\psi\rangle\in\hr_M^{(n)}(\tau)$ into
the $\lceil \log_2 d\rceil$-qubit-space $\left(\C^2\right)^{\otimes \lceil \log_2 d\rceil}$.
Thus, the qubit string ${\mathcal C} |\varphi\rangle$ of length $\lceil \log_2 d\rceil\leq n$
can be considered as the ``quantum part'' of $|\varphi\rangle$.

So how can the classical part of $|\varphi\rangle$ be encoded into a short classical binary string?
Our task is to specify
what halting space $\hr_M^{(n)}(\tau)$ corresponds to $|\varphi\rangle$. Unfortunately, it
is not possible to encode the halting time $\tau$ directly, since $\tau$ might be huge and
may not have a short description. Instead, we can encode the {\em halting number}.
Define the halting time sequence $\{t_i\}_{i=1}^N$ as the set of all integers
$t\in\N$ such that $\dim \hr_M^{(n)}(t)\geq 1$, ordered such that $t_i<t_{i+1}$
for every $i$, that is, the set of all halting times that can occur on inputs of length $n$.
Thus, there must be some $i\in\N$ such that $\tau=t_i$, and $i$ can be called the
halting number of $|\varphi\rangle$. Now, we assign code words $c_i$ to the halting
numbers $i$, that is, we construct a prefix code $\{c_i\}_{i=1}^N\subset\{0,1\}^*$.
We want the code words to be short; we claim that we can always choose the lengths as
\[
   \ell(c_i)=n+1-\lceil \log_2 \dim \hr_M^{(n)}(t_i)\rceil\,\,.
\]
This can be verified by checking the Kraft inequality:
\begin{eqnarray*}
   \sum_{i=1}^N 2^{-\ell(c_i)}&=&2^{-n}\sum_{i=1}^N 2^{\lceil \log_2 \dim \hr_M^{(n)}(t_i)\rceil
   -1}\\
   &\leq& 2^{-n} \sum_{i=1}^n \dim \hr_M^{(n)}(t_i)
   \leq 2^{-n} \dim \left(\C^2\right)^{\otimes n}\\
   &\leq& 1,
\end{eqnarray*}
since the halting spaces are mutually orthogonal.

Putting classical and quantum part of $|\varphi\rangle$ together, we get
\[
   |\tilde\varphi\rangle:=c_i\otimes {\mathcal C} |\varphi\rangle\,\,,
\]
where $i$ is the halting number of $|\varphi\rangle$.
Thus, the length of $|\tilde\varphi\rangle$ is exactly $n+1$.

Let $s_M$ be a self-delimiting description of the QTM $M$. The idea is to construct
a QTM $\mathfrak U$ that, on input $s_M\otimes |\tilde\varphi\rangle$, proceeds as follows:
\begin{itemize}
\item By {\em classical} simulation of $M$, it computes descriptions of the halting spaces
$\hr_M^{(n)}(1), \hr_M^{(n)}(2), \hr_M^{(n)}(3),\ldots$ and the corresponding code words
$c_1,c_2,c_3,\ldots$ one after the other, until at step $\tau$, it finds the code word
$c_i$ that equals the code word in the input.
\item Afterwards, it applies a (quantum) decompression map to approximately reconstruct $|\varphi\rangle$
from ${\mathcal C}|\varphi\rangle$.
\item Finally, it simulates (quantum) for $\tau$ time steps the time evolution of $M$ on
input $|\varphi\rangle$ and then halts, whatever happens with the simulation.
\end{itemize}
Such a QTM $\mathfrak U$ will have the strong universality property as stated in
Theorem~\ref{MainTheorem1}. Unfortunately, there are many difficulties that
have to be overcome by the proof in Section~\ref{SecConstruction}:
\begin{itemize}
\item Also classically, QTMs can only be simulated approximately. Thus, it is for example impossible
for $\mathfrak U$ to decide by classical simulation whether the QTM $M$ halts on some input
$|\psi\rangle$ perfectly or only approximately at some time $t$. Thus, we have to define
certain $\delta$-approximate halting spaces $\hr_M^{(n,\delta)}(t)$ and prove a lot of lemmas
with nasty inequalities.
\item Since our approach includes mixed qubit strings, we have to consider mixed inputs and outputs as well.
\item The aforementioned prefix code must have the property that one code word can be constructed
after the other (since the sequence of all halting times is not computable), see Lemma~\ref{LemBlindCoding}.
\end{itemize}
We show that all these difficulties (and some more) can be overcome, and the idea outlined
above can be converted to a formal proof of Theorem~\ref{MainTheorem1} and the second part
of Theorem~\ref{MainTheorem2} which we give in full detail in Section~\ref{SecConstruction}.

For the first part of Theorem~\ref{MainTheorem2},
concerning the complexity notion $QC$, a more general result is needed which
is stated in Proposition~\ref{PropTwoParameter}, since this complexity notion
needs an additional parameter as input. For this proposition, the proof idea
outlined above needs to be modified.
The idea for the modified proof of that
proposition is to make the QTM $\mathfrak U$ determine the halting number
of the input (and thus the halting time) directly by projective measurement in the basis
of (approximations of) the halting spaces. We will not prove Proposition~\ref{PropTwoParameter}
in full detail, but only sketch the proof there, since the technical details are
similar to that of the proof of Theorem~\ref{MainTheorem1}.

\section{Mathematical Framework and Formalism}
\label{SecFramework}
Here, we introduce the formalism that is used in Section~\ref{SecConstruction}
to describe qubit strings, quantum Turing machines, and quantum Kolmogorov complexity.
We denote the density operators on a Hilbert space $\hr$ by $\mathcal{T}_1^+(\hr)$ (i.e. the
positive trace-class operators with trace $1$). The natural numbers will be denoted
$\N=\{1,2,3,\ldots,\}$, and we use the symbols $\N_0:=\N\cup\{0\}$ and
$\R_0^+:=\{x\in\R\,\,|\,\,x\geq 0\}$ as well as $\delta_{t't}$, which shall be $1$ if $t'=t$ and $0$ otherwise.

\subsection{Indeterminate-Length Qubit Strings}
\label{SubsecQubitStrings}
The quantum analogue of a bit string, a so-called {\em qubit string}, is a superposition of several classical bit strings.
To be as general as possible, we would like to allow also superpositions of strings of {\em different} lengths like
\begin{equation}
   |\varphi\rangle:=\frac 1 {\sqrt 2}\left( |00\rangle+|11011\rangle\right).
   \label{EqVarphi}
\end{equation}
Such quantum states are called {\em indeterminate-length qubit strings}. They have been studied
by Schumacher and Westmoreland \cite{SchumacherWestmoreland}, as well as by Bostr\"om and Felbinger \cite{Bostroem} in the context
of lossless quantum data compression.

Let \label{HK}$\hr_n:=\left(\C^{\{0,1\}}\right)^{\otimes n}$ be the Hilbert space
of $n$ qubits ($n\in\N_0$). We write $\C^{\{0,1\}}$ for $\C^2$ to
indicate that we fix two orthonormal {\em computational basis vectors}
$|0\rangle$ and $|1\rangle$. The Hilbert space $\hr_\s$ which contains indeterminate-length qubit strings like $|\varphi\rangle$
can be formally defined as the direct sum
\[
   \hr_{\s}:=\bigoplus_{k=0}^\infty \hr_k.
   \label{DefHrS}
\]
The classical finite binary strings $\{0,1\}^*$ are identified with
the computational basis
vectors in $\hr_\s$, i.e. $\hr_{\s}\simeq \ell^2(\{\lambda,
0,1,00,01,\ldots\})$, where $\lambda$ denotes the empty string.
We also use the notation $\hr_{\leq n}:=\bigoplus_{k=0}^n \hr_k$
and treat it as a subspace of $\hr_\s$.

To be as general as possible,
we do not only allow superpositions of strings of different lengths,
but also {\em mixtures}, i.e. our qubit strings are arbitrary density operators
on $\hr_\s$. It will become clear in the next sections that QTMs naturally
produce mixed qubit strings as outputs. Moreover, it will be a useful feature
that the result of applying the partial trace to segments of qubit strings
will itself be a qubit string.

\begin{definition}[Qubit Strings and their Length]
\label{DefQubitStringsandtheirLength}
\lineclear
An (indeterminate-length) {\em qubit string} $\sigma$ is a density operator on $\hr_\s$.
Normalized vectors $|\psi\rangle\in\hr_\s$ will also be called qubit strings, identifying them
with the corresponding density operator $|\psi\rangle\langle\psi|$.
The {\em base length} (or just {\em length}) of a qubit string $\sigma\in\mathcal{T}_1^+(\hr_\s)$ is
defined as
\[
   \ell(\sigma):=\max\{\ell(s)\,\,|\,\,\langle s|\sigma|s\rangle>0,\,\, s\in\s\}
\]
or as $\ell(\sigma)=\infty$ if the maximum does not exist.
\end{definition}
For example,
the density operator $|\varphi\rangle\langle\varphi|$ with $|\varphi\rangle$ as defined in Equation~(\ref{EqVarphi})
is a (pure) qubit string of length $\ell(|\varphi\rangle\langle\varphi|)=5$. This corresponds to the fact that this state $|\varphi\rangle$
needs at least $5$ cells on a QTM's tape to be stored perfectly (compare
Subsection~\ref{SubsecQTMs}). An alternative approach would be to consider the expectation value
$\bar\ell$ of the length instead, which has been proposed by
Rogers and Vedral \cite{Rogers}, see also the discussion in Section~\ref{SecConclusions}.

In contrast to classical bit strings, there are uncountably
many qubit strings that cannot be perfectly distinguished by means
of any quantum measurement. A good measure for the difference between two qubit strings $\sigma$ and $\rho$ is
the trace distance (cf. \cite{NielsenChuang})
\begin{equation}
   \|\rho-\sigma\|_{\rm Tr}:=\frac 1 2 {\rm Tr} \left\vert \rho-\sigma
   \right\vert=\frac 1 2 \sum_i |\lambda_i |,
   \label{tr-dist}
\end{equation}
where the $\lambda_i$ are the eigenvalues of the trace-class operator $\rho-\sigma$. Its operational
interpretation is that it gives the maximum probability of correctly distinguishing between $\rho$ and
$\sigma$ by means of any single quantum measurement.

\subsection{Mathematical Description of Quantum Turing Machines}
\label{SubsecQTMs}
Bernstein and Vazirani (\cite{BernsteinVazirani}, Def. 3.2.2) define
a quantum Turing machine $M$ as a triplet $(\Sigma,Q,\delta)$, where
$\Sigma$ is a finite alphabet with an identified blank symbol $\#$, and
$Q$ is a finite set of states with an identified initial state $q_0$
and final state $q_f\neq q_0$. The function $\delta:Q\times\Sigma\to\tilde\C^{\Sigma\times Q \times \{L,R\}}$
is called the {\em quantum transition function}.
The symbol $\tilde\C$ denotes the set of complex numbers $\alpha\in\C$ such that there is
a deterministic algorithm that computes the real and imaginary parts of $\alpha$ to within $2^{-n}$
in time polynomial in $n$.

One can think of a QTM as consisting of
a two-way infinite tape $\mathbf T$ of cells indexed by $\Z$, a control $\mathbf C$, and a single
``read/write'' head $\mathbf H$ that moves along the tape. A QTM evolves in discrete, integer
time steps, where at every step, only a finite number of tape cells is non-blank.
For every QTM, there is a corresponding Hilbert space
$\hr_{QTM}=\hr_{\mathbf{C}}\otimes \hr_{\mathbf{T}}
\otimes \hr_{\mathbf{H}}$, where $\hr_{\mathbf C}=\C^Q$ is a finite-dimensional Hilbert space
spanned by the (orthonormal) control states $q\in Q$, while $\hr_{\mathbf T}=\ell^2(T)$
and $\hr_{\mathbf H}=\ell^2(\mathbb{Z})$
are separable Hilbert spaces describing the contents of the tape
and the position of the head,
where
\begin{equation}
   T=\left\{(x_i)_{i\in\Z}\in\Sigma^\Z\,\,|\,\, x_i\neq\#
   \mbox{ for finitely many }i\in\Z\right\}
   \label{EqDefT}
\end{equation}
denotes the set of classical tape configurations
with finitely many non-blank symbols.

For our purpose, it is useful to consider a special class of QTMs
with the property that their tape $\mathbf T$ consists of two different tracks
(cf. \cite[Def. 3.5.5]{BernsteinVazirani}),
an {\em input track} $\mathbf I$ and an {\em output track} $\mathbf O$. This can be achieved by
having an alphabet which is a Cartesian product of two alphabets,
in our case $\Sigma=\{0,1,\#\}\times \{0,1,\#\}$. Then, the tape Hilbert space
$\hr_{\mathbf T}$ can be written as $\hr_{\mathbf{T}}=\hr_{\mathbf{I}}\otimes
\hr_{\mathbf{O}}$.

The transition
function $\delta$ generates a linear operator $U_M$ on $\hr_{QTM}$ describing the
time evolution of the QTM $M$. 
If $\delta$ is chosen in accordance with certain conditions,
then $U_M$ will be unitary (and thus compatible with quantum theory), see Ozawa and Nishimura~\cite{Ozawa}.
We identify $\sigma\in\mathcal{T}_1^+(\hr_\s)$ with the initial state
of $M$ on input $\sigma$, which is according to the definition in
\cite{BernsteinVazirani} a state on  $\hr_{QTM}$ where $\sigma$ is written on the input track
over the cell interval $[0, \ell(\sigma)-1]$, the empty state
$\#$ is written on the remaining cells of the input track and on the whole
output track, the control is in the initial state $q_0$ and the head is
in position $0$. By linearity, this e.g. means that the vector $|\psi\rangle=\frac 1 {\sqrt 2}
\left(|0\rangle+|11\rangle\right)$ is identified with the vector
$\frac 1 {\sqrt 2}\left(|0\#\rangle+|11\rangle\right)$
on input track cells number $0$ and $1$.

The global state $M^t(\sigma)\in\mathcal{T}_1^+(\hr_{QTM})$
of $M$ on input $\sigma$ at time $t\in\N_0$ is given by $M^t(\sigma)=\left(U_M\right)^t \sigma
\left(U_M^*\right)^t$. The state of the control at time $t$ is thus given by partial trace
over all the other parts of the machine, that is $M_{\mathbf C}^t(\sigma):={\rm Tr}_{\mathbf{T,H}}\left(
M^t(\sigma)\right)$ (similarly for the other parts of the QTM).
In accordance with \cite[Def. 3.5.1]{BernsteinVazirani}, we say that the
QTM $M$ {\em halts at time $t\in\N$ on input $\sigma\in\mathcal{T}_1^+(\hr_\s)$},
if and only if
\begin{equation}
   \langle q_f|M_{\rm\bf C}^t(\sigma)|q_f\rangle=1 \mbox{ and }
   \langle q_f|M_{\rm\bf C}^{t'}(\sigma)|q_f\rangle=0\quad
   \forall t'<t,
   \label{EqHalting}
\end{equation}
where $q_f\in Q$ is the final state of the control (specified in the
definition of $M$) signalling the halting of the computation. See Subsection~\ref{SubsecIntroQTM}
for a detailed discussion of these halting conditions (\ref{EqHalting}).

In this paper, when we talk about a QTM, we do not mean the machine model itself,
but rather refer to the corresponding partial function on the qubit strings
which is computed by the QTM. Note
that this point of view is different from e.g. that of Ozawa~\cite{OzawaLocalTransition}
who describes a QTM as a map from $\Sigma^*$ to the set of probability distributions on $\Sigma^*$.

We still have to define what is meant by the output of a QTM $M$, once it has halted at some time $t$ on some input
qubit string $\sigma$.
We could take the state of the output tape $M_{\mathbf O}^t(\sigma)$ to be the output, but this is not
a qubit string, but instead a density operator on the Hilbert space $\hr_{\mathbf O}$. Hence, we define
a quantum operation $\mathcal{R}$ which maps the density operators on $\hr_{\mathbf O}$ to density operators
on $\hr_\s$, i.e. to the qubit strings. The operation $\mathcal{R}$ ``reads'' the output from the tape.

\begin{definition}[Reading Operation]
\label{DefReadingOp}
\lineclear
A quantum operation $\mathcal{R}:\mathcal{T}(\hr_{\mathbf O})\to\mathcal{T}(\hr_\s)$ is
called a {\em reading operation}, if for every finite set of classical strings $\{s_i\}_{i=1}^N\subset\s$, it holds that
\begin{eqnarray*}
   \mathcal{R}\left(\mathbb{P}\left(
         \sum_{i=1}^N \alpha_i \left|\begin{array}{ccccccc}
               \ldots & \# & \# & s_i & \# & \# & \ldots \\
                & \mbox{\tiny -2} & \mbox{\tiny -1} & \mbox{\tiny 0} & \mbox{\tiny$\ell(s_i)$} & \mbox{\tiny $\ell(s_i)+1$} &
             \end{array}
         \right\rangle
      \right)
   \right)\\
   =\mathbb{P}\left(
      \sum_{i=1}^N \alpha_i |s_i\rangle
   \right)
\end{eqnarray*}
where $\mathbb{P}(|\varphi\rangle):=|\varphi\rangle\langle\varphi|$ denotes the projector onto $|\varphi\rangle$.
\end{definition}

The condition specified above does not determine $\mathcal{R}$ uniquely; there are many different reading operations.
For the remainder of this paper, we fix the reading operation $\mathcal{R}$ which is specified in the following example.

\begin{example}
\label{ExR}
Let $T$ denote the classical output track configurations as defined in Equation~(\ref{EqDefT}), with
$\Sigma=\{0,1,\#\}$. Then, for every $t\in T$, let $R(t)$ be the classical string that consists of
the bits of $T$ from cell number zero to the last non-blank cell, i.e.
\begin{eqnarray*}
   R&:&T\to\s \\
   &&\left(\begin{array}{ccccccc}
               \ldots & ? & ? & s & \# & ? & \ldots \\
                & \mbox{\tiny -2} & \mbox{\tiny -1} & \mbox{\tiny 0} & \mbox{\tiny$\ell(s)$} & \mbox{\tiny $\ell(s)+1$} &
             \end{array}\right)
   \mapsto s.
\end{eqnarray*}
For every $s\in\s$, there is a countably-infinite number of $t\in T$ such that $R(t)=s$. Thus, to every $t\in T$,
we can assign a natural number $n(t)$ which is the number of $t$ in some enumeration of the set $\{t'\in T\,\,|\,\, R(t')=R(t)\}$;
we only demand that $n(t)=1$ if $t=\left(\begin{array}{ccccccc}
               \ldots & \# & \# & s & \# & \# & \ldots \\
                & \mbox{\tiny -2} & \mbox{\tiny -1} & \mbox{\tiny 0} & \mbox{\tiny$\ell(s)$} & \mbox{\tiny $\ell(s)+1$} &
             \end{array}\right)$.
Hence, if (as usual) $\ell^2\equiv \ell^2(\N)$ denotes the Hilbert space of square-summable sequences, then the map $U$,
defined by linear extension of
\begin{eqnarray*}
   U:\hr_{\mathbf O}&\to& \hr_\s\otimes \ell^2 \\
   |t\rangle&\mapsto& |R(t)\rangle\otimes |n(t)\rangle,
\end{eqnarray*}
is unitary. Then, the quantum operation
\begin{eqnarray*}
   \mathcal{R}:\mathcal{T}(\hr_{\mathbf O})&\to&\mathcal{T}(\hr_\s) \\
   \rho&\mapsto& {\rm Tr}_{\ell^2} \left( U\rho U^*\right)
\end{eqnarray*}
is a reading operation.
\end{example}

We are now ready to define QTMs as partial maps on the qubit strings.

\begin{definition}[Quantum Turing Machine (QTM)]
\label{DefQTM}
\lineclear
A partial map $M:\mathcal{T}_1^+(\hr_\s)\to\mathcal{T}_1^+(\hr_\s)$
will be called a QTM, if there is a Bernstein-Vazirani two-track QTM $M'=(\Sigma,Q,\delta)$
(see \cite{BernsteinVazirani}, Def. 3.5.5)
with the following properties:
\begin{itemize}
\item $\Sigma=\{0,1,\#\}\times \{0,1,\#\}$,
\item the corresponding time evolution operator $U_{M'}$ is unitary,
\item if $M'$ halts on input $\sigma$ at some time $t\in\N$, then
$M(\sigma)=\mathcal{R}\left({M'}_{\mathbf O}^t(\sigma)\right)$, where
$\mathcal{R}$ is the reading operation specified in Example~\ref{ExR} above.
Otherwise, $M(\sigma)$ is undefined.
\end{itemize}
A {\em fixed-length QTM} is the restriction of a QTM to the domain
$\bigcup_{n\in\N_0} \mathcal{T}_1^+(\hr_n)$
of length eigenstates.
\end{definition}

The definition of halting, given by Equation~(\ref{EqHalting}), is very important, as explained
in Subsection~\ref{SubsecIntroQTM}. On the other hand, changing certain details in a QTM's definition,
like the way to read the output or allowing a QTM's head to stay at its position instead of
turning left or right, should not change the results in this paper.

\subsection{Quantum Kolmogorov Complexity}
\label{SubsecQComplexity}
Quantum Kolmogorov complexity has first been defined by Berthiaume,
van Dam, and Laplante~\cite{Berthiaume}. They define the complexity $QC(\rho)$ of
a qubit string $\rho$ as the length of the shortest qubit string that, given as input
into a QTM $M$, makes $M$ output $\rho$ and halt. Since there are uncountably many
qubit strings, but a QTM can only apply a countable number of transformations (analogously
to the circuit model), it is necessary to introduce a certain error tolerance $\delta>0$.

This can be done in essentially two ways: First, one can just fix some tolerance $\delta$.
Second, one can demand that the QTM outputs the qubit string $\rho$ as accurately as
one wants, by supplying the machine with a second parameter as input that represents
the desired accuracy. This is analogous to a classical computer program that computes
the number $\pi=3.14\ldots$: A second parameter $k\in\N$ can make the program output
$\pi$ to $k$ digits of accuracy, for example. We consider both approaches and follow
the lines of \cite{Berthiaume} except for two simple modifications: we use the trace
distance rather than the fidelity, and we also allow indeterminate-length and mixed input and
output qubit strings.

\begin{definition}[Quantum Kolmogorov Complexity]
\label{DefQComplexity}
Let $M$ be a QTM and $\rho\in\mathcal{T}_1^+(\hr_\s)$ a qubit string.
For every $\delta>0$, we define the {\em finite-error quantum Kolmogorov complexity}
$QC_M^\delta(\rho)$ as the minimal length of any qubit string $\sigma\in\mathcal{T}_1^+(\hr_\s)$
such that the corresponding output $M(\sigma)$ has trace distance from $\rho$ smaller
than $\delta$,
\[
   QC_M^\delta(\rho):=\min\left\{\ell(\sigma)\,\,|\,\,\|\rho-M(\sigma)\|_{\rm Tr}<\delta\right\}.
\]
Similarly, we define the {\em approximation-scheme quantum Kolmogorov complexity} $QC_M(\rho)$
as the minimal length of any qubit string $\sigma\in\mathcal{T}_1^+(\hr_\s)$ such that when
given $M$ as input together with any integer $k$, the output $M(\sigma,k)$ has trace distance
from $\rho$ smaller than $1/k$:
\[
   QC_M(\rho):=\min\left\{\ell(\sigma)\,\,\left\vert
      \|\rho-M(\sigma,k)\|_{\rm Tr}< \frac 1 k \forall k\in\N
   \right.\right\}.
\]
\end{definition}
For the definition of $QC_M$, we have to fix a map to encode two inputs
(a qubit string and an integer) into one qubit string; this is easy, see e.g. \cite{Vitanyibook} for the
classical case and \cite{QBrudno} for the quantum case. Also, using $f(k):=1/k$ as accuracy required on
input $k$ is not important; any other computable and strictly decreasing function $f$ that tends to zero for $k\to\infty$
such that $f^{-1}$ is also computable will give the same result up to an additive constant.

Note that if $M$ is at least able to move input data to the output track, then
it holds $QC_M^\delta(\rho)\leq \ell(\rho)+c_M$ with some constant $c_M\in\N$ (and similarly
for $QC_M$). In \cite{QBrudno}, we have shown
that for ergodic quantum information sources, emitted states $|\psi\rangle\in \left(\C^2\right)^{\otimes n}$
have a complexity rate $\frac 1 n QC_{\mathcal U}^\bullet (|\psi\rangle)$ that is with
asymptotic probability $1$ arbitrarily close to the von Neumann entropy rate $s$ of the source.
This demonstrates that quantum Kolmogorov complexity is a useful notion, and that
it is feasible to prove interesting theorems on it.

While this complexity notion $QC(\rho)$ counts the length of the shortest qubit string that makes a QTM
output $\rho$ and halt, there have been different definitions for quantum algorithmic
complexity by Vit\'anyi~\cite{Vitanyi} and G\'acs~\cite{Gacs}. Their approaches are based
on classical descriptions and universal density matrices respectively and are not considered
in this paper since they do not have the invariance problem outlined in Subsection~\ref{SubsecInvariance}.

Note also that
Definition~\ref{DefQComplexity} depends on the definition of the length $\ell(\sigma)$
of a qubit string $\sigma\in\mathcal{T}_1^+(\hr_\s)$; there is a different approach by
Rogers and Vedral~\cite{Rogers} that uses the expected (average) length $\bar \ell$ instead
and results in a different notion of quantum Kolmogorov complexity. The results of this paper
are applicable to that definition, too, as long as the notion of halting of the corresponding
quantum computer is defined in a deterministic way as in Equation~(\ref{EqHalting}).

\section{Construction of a Strongly Universal QTM}
\label{SecConstruction}
\subsection{Halting Subspaces and their Orthogonality}
\label{SubsecHaltingSubspaces}
As already explained in Subsection~\ref{SubsecIntroQTM} in the introduction, restricting
to pure input qubit strings $|\psi\rangle\in\hr_n$ of some fixed length $\ell(|\psi\rangle)=n$,
the vectors with equal halting time $t$ form a linear subspace of $\hr_n$. Moreover, inputs
with different halting times are mutually orthogonal, as depicted in Figure~\ref{AbbHaltingSpaces}.
We will now use the formalism for QTMs introduced in Subsection~\ref{SubsecQTMs} to give a formal proof
of these statements. We use the subscripts $\mathbf C$, $\mathbf I$, $\mathbf O$ and $\mathbf H$
to indicate to what part of the tensor product Hilbert space a vector belongs.

\begin{definition}[Halting Qubit Strings]
\label{DefHaltingQubitStrings}
\lineclear
Let $\sigma\in\mathcal{T}_1^+(\hr_\s)$ be a qubit string and $M$ a quantum Turing machine.
Then, $\sigma$ is called {\em $t$-halting (for $M$)}, if
$M$ halts on input $\sigma$ at time $t\in\N$. We define the {\em halting sets} and {\em halting subspaces}
\begin{eqnarray*}
   H_M(t)&:=&\{|\psi\rangle\in\hr_\s\,\, |\,\, |\psi\rangle\langle\psi|\mbox{ is }
   t\mbox{-halting for }M\},\\
   \hr_M(t)&:=&\{\alpha |\psi\rangle\enspace|\enspace |\psi\rangle\in H_M(t),
   \alpha\in\R\}, \\
   H_M^{(n)}(t)&:=&H_M(t)\cap \hr_n,\qquad \hr_M^{(n)}(t):=\hr_M(t)\cap \hr_n.
\end{eqnarray*}
\end{definition}
Note that the only difference between $H_M^{(n)}(t)$ and $\hr_M^{(n)}(t)$ is that the latter
set contains non-normalized vectors. It will be shown below that $\hr_M^{(n)}(t)$ is indeed
a linear subspace.

\begin{theorem}[Halting Subspaces]
\lineclear
For every QTM $M$, $n\in\N_0$ and $t\in\N$, the sets $\hr_M(t)$ and $\hr_M^{(n)}(t)$ are linear subspaces of
$\hr_\s$ resp. $\hr_n$, and
\[
   \hr_M^{(n)}(t) \perp \hr_M^{(n)}(t')\quad\mbox{and}\quad\hr_M(t) \perp \hr_M(t') \quad\mbox{if }t\neq t'.
   \nonumber
\]
\end{theorem}
{\bf Proof.} Let $|\varphi\rangle,|\psi\rangle\in H_M(t)$. The property that $|\varphi\rangle$
is $t$-halting is equivalent to the statement that there are states $|\Phi_q^{t'}\rangle
\in\hr_{\mathbf I}\otimes \hr_{\mathbf O}\otimes\hr_{\mathbf H}$ and coefficients $c_q^{t'}\in\C$
for every $t'\leq t$ and $q\in Q$ such that
\begin{eqnarray}
   V_M^t\left(\strut|\varphi\rangle_{\mathbf I}\otimes |\Psi_0\rangle\right)
   &=&|q_f\rangle_{\mathbf C} \otimes |\Phi_{q_f}^t\rangle\,\,,
   \label{Condition1}\\
   V_M^{t'}\left(\strut|\varphi\rangle_{\mathbf I}\otimes |\Psi_0\rangle\right)
   &=&\sum_{q \neq q_f}c_q^{t'}
   |q\rangle_{\mathbf C} \otimes |\Phi_q^{t'}\rangle\quad\forall t'<t,
   \label{Condition2}
\end{eqnarray}
where $V_M$ is the unitary time evolution operator for the QTM $M$ as a whole, and
$|\Psi_0\rangle=|q_0\rangle_{\mathbf C}\otimes |\#\rangle_{\mathbf O}
\otimes |0\rangle_{\mathbf H}$ denotes the initial state of the control, output track and head.
Note that $|\Psi_0\rangle$ does not depend on the input qubit string (in this case $|\varphi\rangle$).

An analogous equation holds for $|\psi\rangle$, since it is also $t$-halting by assumption.
Consider a normalized superposition $\alpha |\varphi\rangle+\beta |\psi\rangle\in\hr_\s$:
\begin{eqnarray*}
   &&V_M^t\left(\strut\right.\left(\alpha|\varphi\rangle_{\mathbf I}+\beta|\psi\rangle_{\mathbf I}\right)\otimes |\Psi_0\rangle
   \left.\strut\right)\\
   &\qquad&\qquad\qquad=\alpha V_M^t |\varphi\rangle_{\mathbf I}\otimes |\Psi_0\rangle
    +\beta V_M^t |\psi\rangle_{\mathbf I}\otimes |\Psi_0\rangle\\
   &\qquad&\qquad\qquad= \alpha |q_f\rangle_{\mathbf C} \otimes |\Phi_{q_f}^t\rangle
   +\beta |q_f\rangle_{\mathbf C}\otimes |\tilde\Phi_{q_f}^t\rangle\\
   &\qquad&\qquad\qquad=|q_f\rangle_{\mathbf C} \otimes \left(
      \alpha |\Phi_{q_f}^t\rangle+\beta |\tilde\Phi_{q_f}^t\rangle
   \right).
\end{eqnarray*}
Thus, the superposition also satisfies condition (\ref{Condition1}), and, by a similar calculation,
condition (\ref{Condition2}). It follows that
$\alpha |\varphi\rangle+\beta |\psi\rangle$ must also be $t$-halting.
Hence, $\hr_M(t)$ is a linear subspace of $\hr_\s$. As the intersection of linear
subspaces is again a linear subspace, so must be $\hr_M^{(n)}(t)$.

Let now $|\varphi\rangle\in H_M(t)$ and $|\psi\rangle\in H_M(t')$ such that
$t<t'$. Again by Equations~(\ref{Condition1}) and (\ref{Condition2}), it holds
\begin{eqnarray*}
   \langle \varphi |\psi\rangle &=& 
   \left(\strut\,_{\mathbf I}\langle \varphi| \otimes \langle \Psi_0|\right)
    \left(V_M^t\right)^* V_M^t \left(\strut| \psi\rangle_{\mathbf I}\otimes
    |\Psi_0\rangle\right)\\
   &=&\sum_{Q\ni q \neq q_f} c_q^t
   \underbrace{\enspace_{\mathbf{C}}\langle q_f|q\rangle_{\mathbf C}}_0
   \cdot  \langle \Phi_{q_f}^t | \tilde \Phi_q^t \rangle=0\,\,.
\end{eqnarray*}
It follows that $\hr_M(t)\perp \hr_M(t')$, and similarly for $\hr_M^{(n)}(\cdot)\subset \hr_M(\cdot)$.
\qed
The physical interpretation of the preceding theorem is straightforward: By linearity
of the time evolution, superpositions of $t$-halting strings are again $t$-halting,
and strings with different halting times can be perfectly distinguished by observing
their halting time.

\subsection{Approximate Halting Spaces}
\label{SubsecApproxHalting}
The aim of this subsection is to show that the halting spaces of a QTM can be numerically approximated
by a classical algorithm. Thus, we give a step by step construction of such an algorithm, and show analytically
that the approximations it computes are good enough for our purpose. The main result is given in Theorem~\ref{TheProperties}.
Before we state that theorem, we fix some notation.
\begin{definition}[$\eps$-$t$-halting Property]
\label{DefEpsTHalting}
A qubit string $\sigma\in\mathcal{T}_1^+(\hr_\s)$ will be called {\em $\eps$-$t$-halting for $M$} for
some $t\in \N$, $\eps\geq 0$ and $M$ a QTM, if and only if
\[
   \langle q_f | M_{\mathbf{C}}^{t'}(\sigma)|q_f\rangle \left\{
      \begin{array}{ll}
         \leq \eps & \mbox{for }t'<t\,\,,\\
         \geq 1-\eps & \mbox{for }t'=t\,\,.
      \end{array}
   \right.
\]
We denote by $S_n:=\left\{|\psi\rangle\in\cn\enspace | \enspace
\| |\psi\rangle\|=1\right\}$ the unit sphere in $\cn\equiv \left(\C^2\right)^{\otimes n}$, and by
$U_\delta(|\varphi\rangle):=\left\{|\psi\rangle\in\cn\enspace |
\enspace \| |\psi\rangle-|\varphi\rangle\|< \delta\right\}$ an open ball.
The ball $U_\delta(|\varphi\rangle)$ will be called
$\eps$-$t$-halting for $M$ if there is some $|\psi\rangle\in U_\delta(|\varphi\rangle)\cap S_n$
which is $\eps$-$t$-halting for $M$. Moreover, we use the following symbols:
\begin{itemize}
\item ${\rm dist}(S,|\varphi\rangle):=\inf_{s\in S} \|\,|s\rangle-|\varphi\rangle\|$
for any subset $S\subset \cn$ and $|\varphi\rangle\in\cn$,
\item $\cnq:=\left\{|\varphi\rangle\in\cn\,\,|\,\, \langle e_k | \varphi\rangle \in \mathbb{Q}+i\mathbb{Q}
\quad\forall k\right\}$,
where $\{|e_k\rangle\}_{k=1}^{2^n}$ denotes the computational basis vectors of $\cn$,
\item $|\varphi^0\rangle:=\frac{|\varphi\rangle}{\|\,|\varphi\rangle\|}$ for every vector
$|\varphi\rangle\in\cn\setminus\{0\}$.
\end{itemize}
\end{definition}
The set of vectors with rational coordinates, denoted $\cnq$, will in the following be used frequently
as inputs or outputs of algorithms. Such vectors can be symbolically added or multiplied with rational
scalars without any error. Also, given $|a\rangle,|b\rangle\in\cnq$, it is an easy task
to decide unambiguously which vector has larger norm than the other (one can compare the rational numbers
$\|\,|a\rangle\|^2$ and $\|\,|b\rangle\|^2$, for example).

Now we are ready to state the main theorem of this subsection:
\begin{theorem}[Computable Approximate Halting Spaces]
\label{TheProperties}
There is a classical algorithm that, given a classical description of a QTM $M$, integers $n\in\N_0$, $t\in\N$,
and a rational parameter $\delta>0$, computes a description of some subspace $\hr_M^{(n,\delta)}(t)\subset\hr_n$
and a rational number $\eps_M^{(n,\delta)}(t)>0$
with the following properties:
\begin{itemize}
\item {\em Almost-Halting:} If $|\psi\rangle\in H_M^{(n,\delta)}(t)$, then $|\psi\rangle$
is $(20\,\delta)$-$t$-halting for $M$.
\item {\em Approximation:} For every $|\psi\rangle\in H_M^{(n)}(t)$, there is a vector $|\psi^{(\delta)}\rangle
\in H_M^{(n,\delta)}(t)$ which satisfies $\|\,|\psi\rangle-|\psi^{(\delta)}\rangle\|<\frac {11} 2 \delta$.
\item{\em Similarity:} If $\delta,\Delta\in\mathbb{Q}^+$ such that $\delta\leq \frac 1 {80}\,\eps_M^{(n,\Delta)}(t)$,
then for every $|\psi\rangle\in H_M^{(n,\delta)}(t)$ there is a vector $|\psi^{(\Delta)}\rangle
\in H_M^{(n,\Delta)}(t)$ which satisfies
$\|\,|\psi\rangle-|\psi^{(\Delta)}\rangle\|<\frac {11} 2 \Delta$.
\item {\em Almost-Orthogonality:} If $|\psi_t\rangle\in H_M^{(n,\delta)}(t)$ and
$|\psi_{t'}\rangle\in H_M^{(n,\delta)}(t')$ for $t\neq t'$, then it holds that $|\langle \psi_t | \psi_{t'}\rangle|
\leq 4\sqrt{5\delta}$.
\end{itemize}
\end{theorem}
The description of this algorithm (Definition~\ref{DefApproxHalt}) and the proof of this theorem (on page~\pageref{ProofTheProperties})
need some lemmas that show how certain computational steps can be accomplished.

\begin{lemma}[Algorithm for $\eps$-$t$-halting-Property of Balls]
\label{TheAlgBalls}
\lineclear
There exists a (classical) algorithm $B$ which, on input $|\varphi\rangle\in\cnq$,
$\delta,\eps\in\mathbb{Q}^+$, $t\in\N$ and a classical description $s_M\in\{0,1\}^*$ of
a fixed-length QTM $M$, always halts and returns either $0$ or $1$
under the following constraints:
\begin{itemize}
\item If $U_\delta(|\varphi\rangle)$ is not $\eps$-$t$-halting for $M$, then the output must be $0$.
\item If $U_\delta(|\varphi\rangle)$ is $\frac\eps 4$-$t$-halting for $M$, then the output must be $1$.
\end{itemize}
\end{lemma}
{\bf Proof.} The algorithm $B$ computes a set of vectors $\{|\varphi_k\rangle\}_{k=1}^N\subset
\cnq$ such that for every vector $|\psi\rangle\in U_\delta(|\varphi\rangle)
\cap S_n$ there is a $k\in\{1,\ldots,N\}$ such that $\|\,|\varphi_k\rangle-|\psi\rangle\|\leq
\frac 3 {64}\,\eps$, and also vice versa (i.e. ${\rm dist}\left(\strut U_\delta(|\varphi\rangle)\cap
S_n,|\varphi_k\rangle\right)\leq \frac 3 {64}\,\eps$ for every $k$).

For every $k\in\{1,\ldots,N\}$, the algorithm simulates the QTM $M$ on input $|\varphi_k\rangle$
classically for $t$ time steps and computes an approximation $a(t')$ of the quantity
$\langle q_f|M_{\mathbf{C}}^{t'}(|\varphi_k\rangle\langle\varphi_k|)|q_f\rangle$
for every $t'\leq t$, such that
\[
   \left| \strut a(t')-\langle q_f|M_{\mathbf{C}}^{t'}(|\varphi_k\rangle\langle\varphi_k|)|q_f\rangle\right|
   <\frac 3 {32}\,\eps \qquad\mbox{for every }t'\leq t\,\,.
\]
How can this be achieved? Since the number of time steps $t$ is finite, time evolution will
be restricted to a finite subspace $\tilde\hr_{\mathbf{T}}\subset
\hr_{\mathbf{T}}$ corresponding
to a finite number of tape cells, which also restricts the state space of the head
(that points on tape cells) to a finite subspace $\tilde\hr_{\mathbf{H}}$.
Thus, it is possible to give a matrix representation of
the time evolution operator $V_M$ on $\hr_{\mathbf{C}}\otimes \tilde\hr_{\mathbf{T}}
\otimes \tilde\hr_{\mathbf{H}}$, and the expression given above can be numerically calculated
just by matrix multiplication and subsequent numerical computation of the partial trace.

Every $|\varphi_k\rangle$ that satisfies \label{EinfachSo}$|a(t')-\delta_{t't}|\leq \frac 5 8 \, \eps$ for every $t'\leq t$
will be marked as ``approximately halting''. If there is at least one $|\varphi_k\rangle$ that
is approximately halting, $B$ shall halt and output $1$, otherwise it shall halt and output $0$.

To see that this algorithm works as claimed, suppose that $U_\delta(|\varphi\rangle)$ is not
$\eps$-$t$-halting for $M$, so for every $|\tilde \psi\rangle\in U_\delta(|\varphi\rangle)$ there
is some $t'\leq t$ such that $\left| \delta_{t't}-\langle q_f|M_{\mathbf{C}}^{t'}(
|\tilde\psi\rangle\langle\tilde\psi|)|q_f\rangle\right|>\eps$. Also, for every $k\in\{1,\ldots,N\}$,
there is some vector $|\psi\rangle\in U_\delta(|\varphi\rangle)\cap S_n$ with $\|\,|\varphi_k\rangle
-|\psi\rangle\|\leq\frac 3 {64}\,\eps$, so
\begin{eqnarray*}
   \Delta_k&:=&
   \left|\delta_{t't}-\langle q_f|M_{\mathbf{C}}^{t'}(|\varphi_k\rangle\langle\varphi_k|)|q_f\rangle
   \right|\\
   &\geq&
   \left|\delta_{t't}-\langle q_f|M_{\mathbf{C}}^{t'}(|\psi\rangle\langle\psi|)|q_f\rangle
   \right|\\
   &-&\left|\langle q_f|M_{\mathbf{C}}^{t'}(|\psi\rangle\langle\psi|)|q_f\rangle
   -\langle q_f|M_{\mathbf{C}}^{t'}(|\varphi_k^0\rangle\langle\varphi_k^0|)|q_f\rangle
   \right|\\
   &-&\left|\langle q_f|M_{\mathbf{C}}^{t'}(|\varphi_k\rangle\langle\varphi_k|)|q_f\rangle
   -\langle q_f|M_{\mathbf{C}}^{t'}(|\varphi_k^0\rangle\langle\varphi_k^0|)|q_f\rangle
   \right|\\
   &>&\eps-\|\,|\psi\rangle\langle\psi|-|\varphi_k^0\rangle\langle\varphi_k^0|\|_{\rm{Tr}}
   -2\cdot\left|1-\|\,|\varphi_k\rangle\|^2\right|\\
   &\geq& \eps-\|\,|\psi\rangle-|\varphi_k^0\rangle\|-2\left|\strut1-\|\,|\varphi_k\rangle\|\right|
   (1+\|\,|\varphi_k\rangle\|)\\
   &\geq& \eps-\frac 3 {64}\,\eps-\|\,|\varphi_k\rangle-|\varphi_k^0\rangle\|-4\cdot\frac 3 {64}\,\eps
   \geq \frac{23}{32}\eps\,\,,
\end{eqnarray*}
where we have used Lemma~\ref{LemNormInequalities} and Lemma~\ref{LemStability}. Thus, for every $k$ it holds
\begin{eqnarray*}
   \left|\strut a(t')-\delta_{t't}\right|&\geq&\Delta_k
   -\left|\strut \langle q_f| M_{\mathbf{C}}^{t'}(|\varphi_k\rangle\langle \varphi_k|)|q_f\rangle
   -a(t')\right|\\
   &>&\frac{23}{32}\eps-\frac 3 {32}\,\eps=\frac 5 8 \eps\,\,,
\end{eqnarray*}
which makes the algorithm halt and output $0$.

On the other hand, suppose that $U_\delta(|\varphi\rangle)$ is $\frac \eps 4$-$t$-halting for $M$,
i.e. there is some $|\psi\rangle\in U_\delta(|\varphi\rangle)\cap S_n$ which is
$\frac \eps 4$-$t$-halting for $M$. By construction, there is some $k$ such that
$\|\,|\varphi_k\rangle-|\psi\rangle\|\leq \frac 3 {64}\,\eps$. A similar calculation as above yields
$\left|\strut \delta_{t't}-\langle q_f|M_{\mathbf{C}}^{t'}(|\varphi_k\rangle\langle\varphi_k|)
|q_f\rangle\right|\leq \frac{17}{32}\eps$ for every $t'\leq t$, so $\left|\strut a(t')-\delta_{t't}\right|
\leq \frac{17}{32}\eps+\frac 3 {32}\,\eps=\frac 5 8\,\eps$, and
the algorithm outputs $1$.\qed

\begin{lemma}[Algorithm $I$ for Interpolating Subspace]
\label{TheAlgInter}
\lineclear
There exists a (classical) algorithm $I$ which, on input $M,N\in\N$,
$|\tilde\varphi_1\rangle,\ldots,|\tilde\varphi_M\rangle$,
$|\varphi_1\rangle,\ldots,|\varphi_N\rangle\in\cnq$,
$d\in\N$, $\mathbb{Q}^+\ni\Delta>\delta$ and $\mathbb{Q}^+\ni\tilde\Delta>\tilde\delta$, always
halts and returns the description of a pair $(i,\tilde U)$ with $i\in\{0,1\}$ and $\tilde U\subset\cn$ a linear
subspace, under the following constraints:
\begin{itemize}
\item If the output is $(1,\tilde U)$, then $\tilde U\subset\cn$ must be a subspace
of dimension $\dim\tilde U=d$ such that ${\rm dist}(\tilde U,|\varphi_k\rangle)<\Delta$ for
every $k$ and ${\rm dist}(\tilde U,|\tilde\varphi_l\rangle)>\tilde\delta$ for every $l$.
\item If there exists a subspace $U\subset\cn$ of dimension $\dim U=d$ such that ${\rm dist}(U,
|\varphi_k\rangle)\leq\delta$ for every $k$ and ${\rm dist}(U,|\tilde\varphi_l\rangle)\geq\tilde\Delta$
for every $l$, then
the output must be of the\footnotemark \enspace form $(1,\tilde U)$.
\end{itemize}
The description of the subspace $\tilde U$ is a list of linearly independent
vectors $\{|\tilde u_i\rangle\}_{i=1}^d\subset\cnq
\cap \tilde U$.
\end{lemma}
\footnotetext{$\tilde U$ will then be an approximation of $U$.}
{\bf Proof.} Proving this lemma is a routine (but lengthy) exercise. The idea is to construct an
algorithm that looks
for such a subspace by brute force, that is, by discretizing the set of all
subspaces within some (good enough) accuracy. We omit the details.\qed

We proceed by defining {\em approximate halting spaces} as the output of a certain algorithm.
It will turn out that these spaces satisfy all the properties
stated in Theorem~\ref{TheProperties}. Note that the definition
depends on the details of the previously defined algorithms in Lemma~\ref{TheAlgBalls}
and \ref{TheAlgInter}
(for example, there are always different possibilities to compute the necessary discretizations).
Thus, we fix a concrete instance of all those algorithms for the rest of the paper.

\begin{definition}[Approximate Halting Spaces]
\label{DefApproxHalt}
\lineclear
We define\footnote{From a formal point of view, the notation should rather read $\hr_{s_M}^{(n,\delta)}(t)$
instead of $\hr_M^{(n,\delta)}(t)$, since this space depends also on the choice of the classical
description $s_M$ of $M$.} the
$\delta$-approximate halting space $\hr_M^{(n,\delta)}(t)\subset\cn$ and the $\delta$-approximate
halting accuracy $\eps_M^{(n,\delta)}(t)\in\mathbb{Q}$
as the outputs
of the following classical algorithm on input $n,t\in\N$, $0<\delta\in\mathbb{Q}$ and
$s_M\in\{0,1\}^*$, where $s_M$ is a classical description of a fixed-length QTM $M$:
\begin{itemize}
\item[(1)] Let $\eps:=18\,\delta$.
\item[(2)] Compute a covering of $S_n$ of open balls of radius $\delta$, that is,
a set of vectors $\{|\psi_1\rangle,\ldots,|\psi_L\rangle\}\subset\cnq$
($L\in\N$) with $\|\,|\psi_k\rangle\|\in \left( 1-\frac\delta 2,1+\frac \delta 2\right)$
for every $k\in\{1,\ldots,L\}$
such that $S_n\subset\bigcup_{i=1}^L U_\delta(|\psi_i\rangle)$.
\item[(3)] For every $k\in\{1,\ldots,L\}$, compute $B(|\psi_k\rangle,\delta,\eps,t,s_M)$
and $B(|\psi_k\rangle,\delta,18\,\delta,t,s_M)$, where $B$ is
the algorithm for testing the $\eps$-$t$-halting property of balls of Lemma~\ref{TheAlgBalls}.
If the output is $0$ for every $k$, then output $\left(\{0\},\eps\right)$
and halt.
Otherwise set for $\N_0\ni N\leq L$ and $\N_0\ni K\leq L$
\begin{eqnarray*}
   \left\{|\varphi_i\rangle\right\}_{i=1}^N&:=&\left\{
      |\psi_k\rangle\enspace|\enspace
      B(|\psi_k\rangle,\delta,\eps,t,s_M)=1
   \right\},\\
   \left\{|\tilde\varphi_i\rangle\right\}_{i=1}^K&:=&\left\{
      |\psi_k\rangle\enspace|\enspace
      B(|\psi_k\rangle,\delta,18\,\delta,t,s_M)=0
   \right\}.
\end{eqnarray*}
If $N=0$, i.e. if the set $\{|\varphi_i\rangle\}_{i=1}^N$ is empty,
output $\left(\{0\},\eps\right)$ and halt.
\item[(4)] Set $d:=2^n$.
\item[(5)] Let $\Delta:=2\delta$, $\tilde\Delta:=\frac 7 4 \delta$ and $\tilde\delta:=\frac 3 2 \delta$.
Use the algorithm $I$ of Lemma~\ref{TheAlgInter} to search for an interpolating
subspace, i.e., compute $I(K,N,|\tilde\varphi_1\rangle,\ldots,|\tilde\varphi_K\rangle,
|\varphi_1\rangle,\ldots,|\varphi_N\rangle,d,\Delta,\delta,\tilde\Delta,\tilde\delta)$. If
the output of $I$ is $(1,\tilde U)$, output $\left(\tilde U,\eps\right)$ and halt.
\item[(6)] Set $d:=d-1$. If $d\geq 1$, then go back to step (5).
\item[(7)] Set $\eps:=\frac\eps 2$ and go back
to step (3).
\end{itemize}
Moreover, let $H_M^{(n,\delta)}(t):=\hr_M^{(n,\delta)}(t)\cap S_n$.
\end{definition}

The following theorem proves that this definition makes sense:

\begin{theorem}
The algorithm in Definition~\ref{DefApproxHalt} always terminates on any input; thus,
the approximate halting spaces $\hr_M^{(n,\delta)}(t)$ are well-defined.
\end{theorem}
{\bf Proof.} 
Define the function $\eps_{min}:S_n\to\R_0^+$
by $\eps_{min}(|\psi\rangle):=\inf\{{\eps>0}\enspace|\enspace |\psi\rangle \mbox{ is }\eps\mbox{-}
   t\mbox{-halting for }M\}$.
Lemma~\ref{LemNormInequalities}
and \ref{LemStability} yield
\begin{equation}
   \left|\strut \eps_{min}(|\psi_1\rangle)-\eps_{min}(|\psi_2\rangle)\right| \leq
   \|\,|\psi_1\rangle-|\psi_2\rangle\|\,\,,
   \label{eqEpsMin}
\end{equation}
so $\eps_{min}$ is continuous.
For the special case $H_M^{(n)}(t)=\emptyset$, it must thus hold that
$\eps_{min}(S_n):=\min_{|\psi\rangle\in S_n} \eps_{min}(|\psi\rangle)>0$. If the
algorithm has run long enough such that $\eps<\eps_{min}(S_n)$, it must then
be true that $B(|\psi_k\rangle,\delta,\eps,t,s_M)=0$ for every $k\in\{1,\ldots,L\}$,
since all the balls $U_\delta(|\psi_k\rangle)$ are
not $\eps$-$t$-halting. This makes the algorithm halt in step (3).

Now consider the case $H_M^{(n)}(t)\neq\emptyset$. The continuous function
$\eps_{min}$ attains a minimum
on every compact set $\bar U_\delta(|\psi_k\rangle)\cap S_n$, so let $\eps_k:=\min_{|\psi\rangle
\in\bar U_\delta(|\psi_k\rangle)\cap S_n} \eps_{min}(|\psi\rangle)$ ($1\leq k\leq N$).
If $\eps_k=0$ for every $k$, then for every $k$ and $\eps>0$, there is some vector $|\psi\rangle
\in U_\delta(|\psi_k\rangle)\cap S_n$ which is $\eps$-$t$-halting for $M$, so
$B(|\psi_k\rangle,\delta,\eps,t,s_M)=1$ for every $\eps>0$, and so $K=0$ in step (3).
Thus, the algorithm $I$ will by construction find the interpolating subspace $\tilde U=\hr_n$
and cause halting in step (5).

Otherwise,
let $\eps_0:=\min\{\eps_k\enspace|\enspace k\in\{1,\ldots,N\},\eps_k>0\}$. Suppose that
the algorithm has run long enough such that $\eps<\eps_0$. By construction of the algorithm
$B$, if $B(|\psi_k\rangle,\delta,\eps,t,s_M)=1$, it follows that $U_\delta(|\psi_k\rangle)$ is
$\eps$-$t$-halting for $M$, but then, $\eps_k\leq \eps < \eps_0$, so $\eps_k=0$, so there is some
$|\psi\rangle\in \bar U_\delta(|\psi_k\rangle)\cap S_n$ which is $0$-$t$-halting for $M$, so
${\rm dist}(\hr_M^{(n)}(t),|\psi_k\rangle)\leq\delta$.
On the other hand, if
$B(|\psi_k\rangle,\delta,18\,\delta,t,s_M)=0$, it follows that $U_\delta(|\psi_k\rangle)$ is not
$\left(\frac 9 2 \delta\right)$-$t$-halting for $M$.
Thus, ${\rm dist}\left(H_M^{(n)}(t),|\psi_k^0\rangle\right)\geq \frac 9 2 \delta$ according to
(\ref{eqEpsMin}),
so ${\rm dist}(\hr_M^{(n)}(t)\cap S_n,|\psi_k\rangle)>4\delta$, and
by elementary estimations ${\rm dist}(\hr_M^{(n)}(t),|\psi_k\rangle)>\frac 7 4 \delta$.
By definition of the
algorithm $I$, it follows that
$I(K,N,|\tilde\varphi_1\rangle,\ldots,|\tilde\varphi_K\rangle,
|\varphi_1\rangle,\ldots,|\varphi_N\rangle,d,\Delta,\delta,\tilde\Delta,\tilde\delta)=(1,\tilde U)$
for $d:=\dim \hr_M^{(n)}(t)\geq 1$ and some subspace $\tilde U\subset \cn$,
which makes the algorithm halt
in step (5).
\qed

We are now ready to prove Theorem~\ref{TheProperties}, by showing that the approximate
halting spaces defined above indeed satisfy the properties stated in that theorem.

{\bf Proof of Theorem~\ref{TheProperties}.} Assume that\label{ProofTheProperties}
$H_M^{(n,\delta)}(t)\neq\emptyset$.
Let $|\psi\rangle\in H_M^{(n,\delta)}(t)\subset S_n$, and let
$\{|\psi_1\rangle,\ldots,|\psi_L\rangle\}\subset\cn$
be the covering of $S_n$ from the algorithm in Definition~\ref{DefApproxHalt}.
By construction, there is some $k\in\{1,\ldots,L\}$ such that
$|\psi\rangle\in U_\delta(|\psi_k\rangle)$. The subspace $\hr_M^{(n,\delta)}(t)$
is computed in step (5) of the algorithm in Definition~\ref{DefApproxHalt}
via
$I(K,N,|\tilde\varphi_1\rangle,\ldots,|\tilde\varphi_K\rangle,
|\varphi_1\rangle,\ldots,|\varphi_N\rangle,d,\Delta,\delta,\tilde\Delta,\tilde\delta)
=(1,\hr_M^{(n,\delta)}(t))$, and since ${\rm dist}(\hr_M^{(n,\delta)}(t),|\psi_k\rangle)<\delta$,
it follows from the properties of the algorithm $I$ in Lemma~\ref{TheAlgInter} that
$|\psi_k\rangle\neq|\tilde\varphi_l\rangle$ for every $l\in\{1,\ldots,K\}$ in step (3)
of the algorithm. Thus, $B(|\psi_k\rangle,\delta,18\,\delta,t,s_M)=1$, and it follows
from the properties of the algorithm $B$ in Lemma~\ref{TheAlgBalls} that
$U_\delta(|\psi_k\rangle)$ is $(18\,\delta)$-$t$-halting for $M$, so there is some
$|\tilde\psi\rangle\in U_\delta(|\psi_k\rangle)\cap S_n$ which is $(18\,\delta)$-$t$-halting for $M$.
Since $\|\,|\tilde\psi\rangle-|\psi\rangle\|<2\delta$, the almost-halting property
follows from Equation~(\ref{eqEpsMin}).

To prove the approximation property, assume that
$H_M^{(n)}(t)\neq\emptyset$.
Let $|\psi\rangle\in H_M^{(n)}(t)\subset S_n$; again, there is some $j\in\{1,\ldots,L\}$ such that
$|\psi\rangle\in U_\delta(|\psi_j\rangle)$, so $U_\delta(|\psi_j\rangle)$ is $0$-$t$-halting for $M$,
and $B(|\psi_j\rangle,\delta,\eps,t,s_M)=1$ for every $\eps>0$ by definition of the
algorithm $B$. For step (3) of the algorithm in Definition~\ref{DefApproxHalt}, it thus
always holds that $|\psi_j\rangle\in\{|\varphi_i\rangle\}_{i=1}^N$. The output of the
algorithm is computed in step (5) via
$I(K,N,|\tilde\varphi_1\rangle,\ldots,|\tilde\varphi_K\rangle,
|\varphi_1\rangle,\ldots,|\varphi_N\rangle,d,\Delta,\delta,\tilde\Delta,\tilde\delta)
=(1,\hr_M^{(n,\delta)}(t))$.
By definition of $I$, it holds
${\rm dist}(\hr_M^{(n,\delta)}(t),|\psi_j\rangle)<\Delta$,
and by elementary estimations it follows that
${\rm dist}(\hr_M^{(n,\delta)}(t)\cap S_n,|\psi_j\rangle) < \frac \delta 2 + 2 \Delta$,
so there is some $|\psi^{(\delta)}\rangle\in H_M^{(n,\delta)}(t)$
such that $\|\,|\psi^{(\delta)}\rangle-|\psi_j\rangle\| < \frac \delta 2 +2\Delta$.
Since $\|\,|\psi\rangle-|\psi_j\rangle\|\leq\delta$ and $\Delta=2\delta$,
the approximation property follows.

Notice that under the assumptions
given in the statement of the similarity property, it follows from the almost-halting property
that if $|\psi\rangle\in H_M^{(n,\delta)}(t)$, then $|\psi\rangle$ must be
$\frac 1 4 \eps_M^{(n,\Delta)}(t)$-$t$-halting for $M$. Consider the computation
of $\hr_M^{(n,\Delta)}(t)$ by the algorithm in Definition~\ref{DefApproxHalt}.
By construction, it always holds that the parameter $\eps$ during the computation
satisfies $\eps\geq \eps_M^{(n,\Delta)}(t)$, so $|\psi\rangle$ is always
$\frac\eps 4$-$t$-halting for $M$, and if $|\psi\rangle\in U_\delta(|\psi_j\rangle)$,
it follows that $B(|\psi_j\rangle,\delta,\eps,t,s_M)=1$. The rest follows in complete
analogy to the proof of the approximation property.

For the almost-orthogonality property, suppose $|v\rangle\in H_M^{(n,\delta)}(t')$ and $|w\rangle\in
H_M^{(n,\delta)}(t)$
are two arbitrary qubit strings of length $n$ with different approximate halting times $t<t'\in\N$.
There is some $l\in\{1,\ldots,L\}$ such that $|w\rangle\in U_\delta(|\psi_l\rangle)$,
so ${\rm dist}(\hr_M^{(n,\delta)}(t),|\psi_l\rangle)<\delta<\tilde\delta$.
Since $I(K,N,|\tilde\varphi_1\rangle,\ldots,|\tilde\varphi_K\rangle,
|\varphi_1\rangle,\ldots,|\varphi_N\rangle,d,\Delta,\delta,\tilde\Delta,\tilde\delta)
=(1,\hr_M^{(n,\delta)}(t))$ at step (5) of the computation of $\hr_M^{(n,\delta)}(t)$,
it follows from the definition of $I$ that there is no $m\in\N$ such that
$|\psi_l\rangle=|\tilde \varphi_m\rangle$ for the sets defined in step (3) of the
algorithm above. Thus, $B(|\psi_l\rangle,\delta,18\,\delta,t,s_M)=1$, and by definition of $B$
it follows that $U_\delta(|\psi_l\rangle)$ must be $(18\,\delta)$-$t$-halting for $M$, so there
is some vector $|\tilde w\rangle\in U_\delta(\psi_l\rangle)\cap S_n$ which is $(18\,\delta)$-$t$-halting for $M$
and satisfies $\|\,|w\rangle-|\tilde w\rangle\|\leq \|\,|\tilde w\rangle-|\psi_l\rangle\|+
\|\,|\psi_l\rangle-|w\rangle\|<2\delta$. Analogously, there is some vector $|\tilde v\rangle\in S_n$
which is $(18\,\delta)$-$t'$-halting for $M$ and satisfies $\|\,|v\rangle-|\tilde v\rangle\|<2\delta$.

From the definition of the trace distance for pure states (see \cite[(9.99)]{NielsenChuang}
and of the $\eps$-$t$-halting property in Definition~\ref{DefEpsTHalting} together
with Lemma~\ref{LemNormInequalities} and Lemma~\ref{LemStability}, it follows that
\begin{eqnarray}
   \sqrt{1-\left| \langle w | v \rangle \right|^2}&=&\|\,|w\rangle\langle w | - |v\rangle\langle
   v|\,\|_{\rm Tr}\nonumber\\
   &\geq& \|\, |\tilde w\rangle\langle\tilde w|-|\tilde v\rangle\langle\tilde v|\,\|_{\rm Tr}\nonumber\\
   &&-\|\,|w\rangle\langle w|-|\tilde w\rangle\langle\tilde w\|\,\|_{\rm Tr}\nonumber\\
   &&-\,\|\,|v\rangle\langle v|-|\tilde v\rangle\langle \tilde v|\,\|_{\rm Tr}\nonumber\\
   &\geq& \left| \langle q_f| M_{\mathbf{C}}^t (|\tilde w\rangle\langle \tilde w|)|q_f\rangle\right.\nonumber\\
   &&\left.\quad-\langle q_f| M_{\mathbf{C}}^t(|\tilde v\rangle\langle\tilde v|)|q_f\rangle\right|\nonumber\\
   &&-\|\,|w\rangle-|\tilde w\rangle\|-\|\,|v\rangle-|\tilde v\rangle\|\nonumber\\
   &\geq& 1-36\,\delta-2\delta-2\delta=1-40\,\delta.
   \label{eqInnerProdTr}
\end{eqnarray}
This proves the almost-orthogonality property.
\qed

The following corollary proves that the approximate halting spaces $\hr_M^{(n,\delta)}(t)$ are ``not too large''
if $\delta$ is small enough.

\begin{corollary}[Dimension Bound for Halting Spaces]
\lineclear
\label{CorDimBound}
If $\delta<\frac 1 {80}\, 2^{-2n}$, then $\displaystyle\sum_{t\in\N} \dim \hr_M^{(n,\delta)}(t)\leq 2^n$.
\end{corollary}
{\bf Proof.} Suppose that $\sum_{t\in\N} \dim \hr_M^{(n,\delta)}(t)> 2^n$. Then, choose orthonormal bases
in each of the spaces $\hr_M^{(n,\delta)}(t)$, and let $\left\{
|\varphi_i\rangle\right\}_{i=1}^{2^n+1}$ be the union of the first $2^n+1$ of these basis vectors.
By construction and by the almost-orthogonality property of Theorem~\ref{TheProperties},
it follows that $|\langle \varphi_i|\varphi_j\rangle|\leq 4\sqrt{5\delta}<2^{-n}=\frac 1
{(2^n+1)-1}$ for every $i\neq j$. Lemma~\ref{LemInnerProduct} yields $\dim U\geq 2^n+1$ for
$U:={\rm span}\left\{|\varphi_i\rangle\right\}_{i=1}^{2^n+1}\subset \cn$, but
$\dim \cn=2^n$, which is a contradiction.\qed

\subsection{Compression, Decompression, and Coding}
\label{SubsecCompDecomp}
In this subsection, we define some compression and coding algorithms that will
be used in the construction of the strongly universal QTM.
\begin{definition}[Standard (De-)Compression]
\label{DefStandard}
\lineclear
Let $U\subset \hr_n$ be a linear subspace with $\dim U=N$. Let
$P_U\in\mathcal{B}(\hr_n)$ be the orthogonal projector onto $U$, and
let $\left\{|e_i\rangle\right\}_{i=1}^{2^n}$ be the computational
basis of $\hr_n$.
The result of applying the Gram-Schmidt orthonormalization procedure to the vectors
$\left\{|\tilde u_i\rangle\right\}_{i=1}^{2^n}=\left\{P_U |e_i\rangle\right\}_{i=1}^{2^n}$ (dropping every null vector)
is called the {\em standard basis} $\{|u_1\rangle,\ldots,|u_N\rangle\}$ of $U$.
Let $|f_i\rangle$ be the $i$-th computational basis vector of $\hr_{\lceil \log N\rceil}$.
The {\em standard compression}
$\mathcal{C}_U:U\to\hr_{\lceil \log N\rceil}$ is then defined by linear extension of
$\mathcal{C}_U(|u_i\rangle):=|f_i\rangle$ for $1\leq i \leq N$,
that is, $\mathcal{C}_U$ isometrically embeds $U$ into $\hr_{\lceil \log N\rceil}$.
A linear isometric map $\mathcal{D}_U:\hr_{\lceil \log N\rceil}\to \hr_n$ will be
called a {\em standard decompression} if it holds that
\[
   \mathcal{D}_U\circ \mathcal{C}_U=\idn_U\,\,.
\]
\end{definition}
It is clear that there exists a classical algorithm that, given a description of $U$ (e.g. a list
of basis vectors $\{|u_i\rangle\}_{i=1}^{\dim U}\subset\cnq$), can effectively compute
(classically) an approximate description of the standard basis of $U$. Moreover,
a quantum Turing machine can effectively apply a standard decompression map to its input:

\begin{lemma}[Q-Standard Decompression Algorithm]
\label{LemStandardDecomp}
\lineclear
There is a QTM $\mathfrak{D}$ which, given a description\footnotemark of a
subspace $U\subset \hr_n$, the integer $n\in\N$, some $\delta\in\mathbb{Q}^+$, and a quantum state $|\psi\rangle\in
\hr_{\lceil \log\dim U\rceil}$, outputs some
state $|\varphi\rangle\in\hr_n$ with the property that $\|\,|\varphi\rangle-\mathcal{D}_U |\psi\rangle\|<\delta$,
where $\mathcal{D}_U$ is some standard decompression map.
\end{lemma}
\footnotetext{(a list of linearly independent vectors $\{|\tilde u_1\rangle,\ldots,|\tilde u_{\dim U}\rangle\}\subset U\cap \cnq$)}
{\bf Proof.} Consider the map $A:\hr_{\lceil \log\dim U\rceil}\to\hr_n$, given by
$A|v\rangle:=|0\rangle^{\otimes (n-\lceil \log\dim U\rceil)}
\otimes |v\rangle$. The map $A$ prepends zeroes to a vector; it maps the computational basis
vectors of $\hr_{\lceil \log\dim U\rceil}$ to the lexicographically first computational basis vectors
of $\hr_n$.
The QTM $\mathfrak{D}$ starts by applying this map $A$ to the input state $|\psi\rangle$ by
prepending zeroes on its tape,
creating a state $|\tilde\psi\rangle:=|0\rangle^{\otimes (n-\lceil \log\dim U\rceil)}
\otimes |\psi\rangle\in\hr_n$.

Afterwards, it applies (classically) the Gram-Schmidt orthonormalization procedure
to the list of vectors $\{|\tilde u_1\rangle,\ldots,|\tilde u_{\dim U}\rangle,
|e_1\rangle,\ldots,|e_{2^n}\rangle\}\subset \cnq$, where the vectors $\{|\tilde u_i\rangle\}_{i=1}^{\dim U}$
are the basis vectors of $U$ given in the input, and
the vectors $\{|e_i\rangle\}_{i=1}^{2^n}$
are the computational basis vectors of $\hr_n$. Since every vector has rational entries
(i.e. is an element of $\cnq$), the Gram-Schmidt procedure can be applied exactly,
resulting in a list
$\{|u_i\rangle\}_{i=1}^{2^n}$ of basis vectors of $\hr_n$ which have entries that are
square roots of rational numbers. Note that by construction,
the vectors $\{|u_i\rangle\}_{i=1}^{\dim U}$ are the standard basis vectors of $U$
that have been defined in Definition~\ref{DefStandard}.

Let $V$ be the unitary $2^n\times 2^n$-matrix that has the vectors $\{|u_i\rangle\}_{i=1}^{2^n}$
as its column vectors.
The algorithm continues by computing a rational approximation $\tilde V$ of $V$ such that the
entries satisfy
$|\tilde V_{ij}-V_{ij}|<\frac\delta {2^{n+1} (10\sqrt{2^n})^{2^n}}$, and thus,
in operator norm, it holds $\|\tilde V-V\|<\frac\delta{2(10\sqrt{2^n})^{2^n}}$.
Bernstein and Vazirani \cite[Sec. 6]{BernsteinVazirani} have shown that there are QTMs that
can carry out an $\eps$-approximation of a desired unitary
transformation $V$ on their tapes if given a matrix $\tilde V$ as input
that is within distance $\frac\eps {2(10\sqrt d)^d}$ of the $d\times d$-matrix $V$.
This is exactly the case here\footnote{Note that we consider $\hr_n$ as a subspace
of an $n$-cell tape segment Hilbert space
$\left(\C^{\{0,1,\#\}}\right)^{\otimes n}$, and we demand $V$ to leave blanks $|\#\rangle$
invariant.}, with $d=2^n$ and $\eps=\delta$, so let the $\mathfrak{D}$ apply $V$ within
$\delta$ on its tape to create the state $|\varphi\rangle\in\hr_n$ with
$\|\,|\varphi\rangle-V|\tilde\psi\rangle\|=\|\,|\varphi\rangle-V\circ A |\psi\rangle\|
<\delta$. Note that the map $V\circ A$
is a standard decompression map (as defined in Definition~\ref{DefStandard}), since
for every $i\in\{1,\ldots,\dim U\}$ it holds that
\[
   V\circ A\circ\mathcal{C}_U |u_i\rangle=V\circ A |f_i\rangle
   =V |e_i\rangle=|u_i\rangle\,\,,
\]
where the vectors $|f_i\rangle$ are the computational basis vectors of $\hr_{\lceil
\log\dim U\rceil}$.\qed

The next lemma will be useful for coding the ``classical part'' of a halting qubit string.
The ``which subspace'' information
will be coded into a classical string $c_i\in\s$ whose length $\ell_i\in\N_0$ depends on the dimension of the
corresponding halting space $\hr_M^{(n,\delta)}(t_i)$. The dimensions of the halting spaces
$\left(\dim \hr_M^{(n,\delta)}(t_1),\dim \hr_M^{(n,\delta)}(t_2),\ldots\right)$ can be computed one after the other,
but the complete list of the code word lengths $\ell_i$ is not computable due to the
undecidability of the halting problem.
Since most well-known prefix codes (like Huffman code, see \cite{CoverThomas}) start by initially sorting
the code word lengths in decreasing order, and thus require complete knowledge of the
whole list of code word lengths in advance, they are not suitable for our purpose. We thus give an easy algorithm
that constructs the code words one after the other, such that code word $c_i$ depends
only on the previously given lengths $\ell_1,\ell_2,\ldots,\ell_i$. We call this
``blind prefix coding'', because code words are assigned sequentially without looking at what is coming next.

\begin{lemma}[Blind Prefix Coding]
\label{LemBlindCoding}
\lineclear
Let $\{\ell_i\}_{i=1}^N\subset \N_0$ be a sequence of natural numbers (code word lengths)
that satisfies the Kraft inequality $\displaystyle\sum_{i=1}^N 2^{-\ell_i}\leq 1$.
Then the following (``blind prefix coding'') algorithm produces a list
of code words $\{c_i\}_{i=1}^N\subset \s$ with $\ell(c_i)=\ell_i$, such that
the $i$-th code word only depends on $\ell_i$ and the previously chosen codewords $c_1,\ldots,c_{i-1}$:
\begin{itemize}
\item Start with $c_1:=0^{\ell_1}$, i.e. $c_1$ is the string consisting of $\ell_1$ zeroes;
\item for $i=2,\ldots,N$ recursively, let $c_i$ be the first string in lexicographical order
of length $\ell(c_i)=\ell_i$ that is no prefix or extension of any of the previously
assigned code words $c_1,\ldots,c_{i-1}$.
\end{itemize}
\end{lemma}

{\bf Proof.} We omit the lengthy, but simple proof; it is based on identifying the binary
code words with subintervals of $[0,1)$ as explained in \cite{Vitanyibook}.
We also remark that the content of this lemma is given in \cite[Thm. 5.2.1]{CoverThomas}
without proof as an example for a prefix code.
\qed

\subsection{Proof of the Strong Universality Property}
\label{SubsecProofMainTheorems}
To simplify the proof of Main Theorem~\ref{MainTheorem1}, we show now that
it is sufficient to consider fixed-length QTMs only:

\begin{lemma}[Fixed-Length QTMs are Sufficient]
\label{LemFixedLengthEnough}
\lineclear
For every QTM $M$, there is a fixed-length QTM $\tilde M$ such that
for every $\rho\in\mathcal{T}_1^+(\hr_\s)$ there is a fixed-length qubit string
$\tilde\rho\in\bigcup_{n\in\N_0} \mathcal{T}_1^+(\hr_n)$ such that
$M(\rho)=\tilde M(\tilde\rho)$ and $\ell(\tilde\rho)\leq\ell(\rho)+1$.
\end{lemma}
{\bf Proof.} Since $\dim \hr_{\leq n}=2^{n+1}-1$, there is an isometric
embedding of $\hr_{\leq n}$ into $\hr_{n+1}$. One example is the map $V_n$,
which is defined as $V_n|e_i\rangle:=|f_i\rangle$ for $i\in\{1,\ldots,2^{n+1}-1\}$,
where $|e_i\rangle$ and $|f_i\rangle$ denote the computational basis vectors
(in lexicographical order) of $\hr_{\leq n}$ and $\hr_{n+1}$ respectively.
As $\hr_{n+1}\subset\hr_{\leq(n+1)}$ and $\hr_{\leq n}\subset \hr_{\leq(n+1)}$, we can extend $V_n$ to a unitary
transformation $U_n$ on $\hr_{\leq(n+1)}$, mapping computational basis vectors
to computational basis vectors.

The fixed-length QTM $\tilde M$ works as follows, given some fixed-length
qubit string $\tilde\rho\in\mathcal{T}_1^+(\hr_{n+1})$ on its input tape: first, it determines
$n+1=\ell(\tilde\rho)$
by detecting the first blank symbol $\#$. Afterwards, it computes a description
of the unitary transformation $U_n^*$ and applies it to the qubit string $\tilde\rho$
by permuting the computational basis vectors in the $(n+1)$-block of cells
corresponding to the Hilbert space $\left(\C^{\{0,1,\#\}}\right)^{\otimes(n+1)}$.
Finally, it calls the QTM $M$ to continue the computation on input 
$\rho:=U_n^*\,\tilde\rho\, U_n$. If $M$ halts, then the output will be $M(\rho$).
\qed

\vskip 0.5cm

{\bf Proof of Theorem~\ref{MainTheorem1}.} First, we show
how the input $\sigma_M$ for the strongly universal QTM $\mathfrak U$ is constructed
from the input $\sigma$ for $M$.
Fix some QTM $M$ and input length $n\in\N_0$, and let
$\eps_0:=\frac 1 {81}\, 2^{-2n}$.
Define the
halting time sequence $\{t_M^{(n)}(i)\}_{i=1}^N$ as the set of all integers $t\in\N$ such
that $\dim \hr_M^{(n,\eps_0)}(t)\geq 1$,
ordered such that $t_M^{(n)}(i)<t_M^{(n)}(i+1)$
for every $i$. The number $N$ is in general not computable, but must be somewhere between $0$
and $2^n$ due to Corollary~\ref{CorDimBound}.

For every $i\in\{1,\ldots,N\}$, define the code word length $\ell_i^{(M,n)}$ as
\[
   \ell_i^{(M,n)}:=n+1-\left\lceil \log\dim \hr_M^{(n,\eps_0)}\left(t_M^{(n)}(i)\right)\right\rceil\,\,.
\]
This sequence of code word lengths satisfies the Kraft inequality:
\begin{eqnarray*}
   \sum_{i=1}^N 2^{-\ell_i^{(M,n)}}&=&2^{-n}\sum_{i=1}^N 2^{\left\lceil \log\dim \hr_M^{(n,\eps_0)}\left(t_M^{(n)}(i)\right)\right\rceil
   -1}\\
   &\leq& 2^{-n}\sum_{i=1}^N \dim\hr_M^{(n,\eps_0)}\left(t_M^{(n)}(i)\right)\\
   &=&2^{-n}\sum_{t\in\N}\dim\hr_M^{(n,\eps_0)}(t)\leq 1\,\,,
\end{eqnarray*}
where in the last inequality, Corollary~\ref{CorDimBound} has been used. Let
$\left\{c_i^{(M,n)}\right\}_{i=1}^N
\subset\s$ be the blind prefix code corresponding to the sequence
$\left\{\ell_i^{(M,n)}\right\}_{i=1}^N$
which has been constructed in Lemma~\ref{LemBlindCoding}.

In the following, we use the space $\hr_M^{(n,\eps_0)}(t)$ as some kind of ``reference space''
i.e. we construct our QTM $\mathfrak U$ such that it expects the standard compression
of states $|\psi\rangle\in\hr_M^{(n,\eps_0)}(t)$ as part of the input. If the desired
accuracy parameter $\delta$ is smaller than $\eps_0$, then some ``fine-tuning'' must take place,
unitarily mapping the state $|\psi\rangle\in\hr_M^{(n,\eps_0)}(t)$ into halting spaces
of smaller accuracy parameter. In the next paragraph, these unitary transformations are constructed.

Recursively, for $k\in\N$, define $\eps_k:=\frac 1 {80} \eps_M^{(n,\eps_{k-1})}(t)$.
Since $\eps_M^{(n,\delta)}(t)\leq 18\delta$ by construction of the algorithm in Definition~\ref{DefApproxHalt},
we have $\eps_k\leq \left(\frac {18}{80}\right)^k\cdot \eps_0\stackrel{k\to\infty}\longrightarrow 0$.
It follows from the approximation property of Theorem~\ref{TheProperties} together with
Lemma~\ref{LemDimBoundSimilar} that $\dim\hr_M^{(n,\eps_k)}(t)\geq \dim \hr_M^{(n)}(t)$.
The similarity property and Lemma~\ref{LemDimBoundSimilar} tell us that
$\dim \hr_M^{(n,\eps_{k-1})}(t)\geq \dim \hr_M^{(n,\eps_{k})}(t)$ for every $k\in\N$,
and there exist isometries $U_k:\hr_M^{(n,\eps_{k})}(t)\to\hr_M^{(n,\eps_{k-1})}(t)$
that, for $k$ large enough, satisfy
\begin{equation}
\|U_k-\idn\|<\frac 8 3 \sqrt{\frac {11} 2 \eps_{k-1}} \left(\frac 5 2\right)^{2^n}
\leq {\rm const}_n \cdot \left(\frac{18}{80}\right)^{\frac k 2}.\label{EqConstn}
\end{equation}
Let now $d:=\lim_{k\to\infty} \dim \hr_M^{(n,\varepsilon_k)}(t)$ and
$c:=\min\left\{ k\in\N\,\,|\,\, \dim \hr_M^{(n,\varepsilon_k)}(t)=d\right\}$.
For any choice of the transformations $U_k$ (they are not unique), let
\[
   \tilde\hr_M^{(n,\eps_k)}(t):=\left\{
      \begin{array}{cl}
         U_{k+1}U_{k+2}\ldots U_c \hr_M^{(n,\eps_c)}(t) & \mbox{if } k<c\,\,,\\
         \hr_M^{(n,\eps_k)}(t) & \mbox{if }k\geq c\,\,.
      \end{array}
   \right.
\]
It follows that the spaces $\tilde\hr_M^{(n,\eps_k)}(t)$ all have the same dimension for every $k\in\N_0$,
and that $\tilde\hr_M^{(n,\eps_k)}(t)\subset\hr_M^{(n,\eps_k)}(t)$.
Define the unitary operators
$\tilde U_k:=U_k\upharpoonright \tilde\hr_M^{(n,\eps_k)}(t)$,
then $\|\tilde U_k^*-\idn\|\leq\|U_k-\idn\|$,
and so the sum $\sum_{k=1}^\infty \|\tilde U_k^*-\idn\|$ converges.
Due to Lemma~\ref{LemCompositionUnitary}, the product $U:=\prod_{k=1}^\infty \tilde U_k^*$ converges
to an isometry $U:\tilde\hr_M^{(n,\eps_0)}(t)\to\hr_n$.
It follows from the approximation property in Theorem~\ref{TheProperties} that\label{rmran}
$\hr_M^{(n)}(t)\subset{\rm ran}(U)$, so we can define a unitary map $U^{-1}:{\rm ran}(U)\to \tilde
\hr_M^{(n,\varepsilon_0)}(t)$ by $U^{-1}(Ux):=x$, and $\hr_M^{(n)}(t)\subset {\rm dom}(U^{-1})$.

Due to Lemma~\ref{LemFixedLengthEnough}, it is sufficient to consider
fixed-length QTMs $M$ only, so we can assume that our input $\sigma$ is a fixed-length
qubit string. Suppose $M(\sigma)$ is defined, and let $\tau\in\N$ be the corresponding
halting time for $M$. Assume for the moment that $\sigma=|\psi\rangle\langle\psi|$
is a pure state, so $|\psi\rangle\in H_M^{(n)}(\tau)$.
Recall the definition of the halting time sequence; it follows that there is some
$i\in\N$ such that $\tau=t_M^{(n)}(i)$. Let
\[
   |\psi^{(M,n)}\rangle:=|c_i^{(M,n)}\rangle\otimes
   \mathcal{C}_{\hr_M^{(n,\eps_0)}(\tau)}U^{-1}|\psi\rangle\,\,,
\]
that is, the blind prefix code of the halting number $i$, followed by
the standard compression (as constructed in Definition~\ref{DefStandard})
of some approximation $U^{-1}|\psi\rangle$ of $|\psi\rangle$ that
is in the subspace $\hr_M^{(n,\eps_0)}(\tau)$.
Note that
\begin{eqnarray*}
   \ell\left(|\psi^{(M,n)}\rangle\right)&=&\ell\left(
   c_i^{(M,n)}\right)+\ell\left(\mathcal{C}_{\hr_M^{(n,\eps_0)}(\tau)}U^{-1}|\psi\rangle\right)\\
   &=&\ell_i^{(M,n)}+\left\lceil \log\dim\hr_M^{(n,\eps_0)}(\tau)\right\rceil=n+1\,\,.
\end{eqnarray*}
If $\sigma=\sum_k \lambda_k |\psi_k\rangle\langle\psi_k|$ is a mixed fixed-length qubit string
which is $\tau$-halting for $M$, every convex component $|\psi_k\rangle$ must also be
$\tau$-halting for $M$, and it makes sense to define
$\sigma^{(M,n)}:=\sum_k\lambda_k |\psi_k^{(M,n)}\rangle
\langle\psi_k^{(M,n)}|$, where every $|\psi_k^{(M,n)}\rangle$
(and thus $\sigma^{(M,n)}$) starts with the same classical code word $c_i^{(M,n)}$,
and still $\sigma^{(M,n)}\in\mathcal{T}_1^+(\hr_{n+1})$.

The strongly universal QTM $\mathfrak U$ expects input of the form
\begin{equation}
   \left(s_M\otimes \sigma^{(M,n)},\delta\right)=:\left(\sigma_M,\delta\right)\,\,,
   \label{EqExpected}
\end{equation}
where $s_M\in\s$ is a self-delimiting description of the QTM $M$.
We will now give a description of how $\mathfrak U$ works; meanwhile, we will
always assume that the input is of the expected form (\ref{EqExpected}) and
also that the input $\sigma$ is a {\em pure} qubit string $|\psi\rangle\langle\psi|$ (we discuss the case
of mixed input qubit strings $\sigma$ afterwards):
\begin{itemize}
\item Read the parameter $\delta$ and the description $s_M$.
\item Look for the first blank symbol $\#$ on the tape to determine the length
$\ell(\sigma^{(M,n)})=n+1$.
\item Compute the halting time $\tau$. This is achieved as follows:
\begin{itemize}
\item[(1)] Set $t:=1$ and $i:=0$.
\item[(2)] Compute a description of $\hr_M^{(n,\eps_0)}(t)$. If $\dim \hr_M^{(n,\eps_0)}(t)=0$,
then go to step (5).
\item[(3)] Set $i:=i+1$ and set
$\ell_i^{(M,n)}:=n+1-\left\lceil \log\dim \hr_M^{(n,\eps_0)}\left(t\right)\right\rceil$.
From the previously computed code word lengths $\ell_j^{(M,n)}$ ($1\leq j\leq i$),
compute the corresponding blind prefix code word $c_i^{(M,n)}$. Bit by bit, compare
the code word $c_i^{(M,n)}$ with the prefix of $\sigma^{(M,n)}$. As soon as any difference is
detected, go to step (5).
\item[(4)] The halting time is $\tau:=t$. Exit.
\item[(5)] Set $t:=t+1$ and go back to step (2).
\end{itemize}
\item Let $|\tilde\psi\rangle$ be the rest of the input, i.e. $\sigma^{(M,n)}=:|c_i^{(M,n)}\rangle
\langle c_i^{(M,n)}|
\otimes |\tilde\psi\rangle\langle\tilde\psi|$ (thus
$|\tilde\psi\rangle=e^{i\theta}\mathcal{C}_{\hr_M^{(n,\eps_0)}(\tau)}U^{-1}|\psi\rangle$ with some
irrelevant phase $\theta\in\R$). Apply
the quantum standard decompression algorithm $\mathfrak D$ given in Lemma~\ref{LemStandardDecomp},
i.e. compute $|\tilde\varphi\rangle:=\mathfrak{D}\left(\hr_M^{(n,\eps_0)}(\tau),
n,\frac\delta 3,|\tilde\psi\rangle\right)$.
Then,
\[
   \left\| \,|\tilde\varphi\rangle-\mathcal{D}_{\hr_M^{(n,\eps_0)}(\tau)}|\tilde\psi\rangle\right\|
   =\left\|\,|\tilde\varphi\rangle-U^{-1}|\psi\rangle\right\|<\frac\delta 3\,\,.
\]
\item Compute an approximation $V:\hr_n\to\hr_n$ of a unitary extension of $U$
with $\left\|U-V\upharpoonright\tilde\hr_M^{(n,\eps_0)}(\tau)\right\|<
\frac{\delta/3}{2(10\sqrt{2^n})^{2^n}}=:\eps$, where $U$ is some ``fine-tuning map'' as
constructed above.
This can be achieved as follows:
\begin{itemize}
\item Choose $N\in\N$ large enough such that $\sum_{k=N+1}^\infty {\rm const}_n\cdot
\left(\frac {18}{80}\right)^{\frac k 2}<\frac \varepsilon 2$, where ${\rm const}_n\in\R$
is the constant defined in Equation~(\ref{EqConstn}).
\item For every $k\in\{1,\ldots,N\}$, find matrices $V_k:\hr_n\to\hr_n$ that
approximate the forementioned\footnote{The isometries $U_k$ are not unique, so they can be chosen 
arbitrarily, except for the requirement that Equation~(\ref{EqConstn}) is satisfied,
and that every $U_k$ depends only on $\hr_M^{(n,\eps_k)}(t)$ and $\hr_M^{(n,\eps_{k-1})}(t)$
and not on other parameters.
} isometries $U_k:\hr_M^{(n,\eps_k)}(t)\to\hr_M^{(n,\eps_{k-1})}(t)$ such that
\[
   \left\| \prod_{k=1}^N \tilde U_k^* - \prod_{k=1}^N V_k^* \upharpoonright
   \tilde\hr_M^{(n,\eps_0)}(t)\right\| < \frac \eps 2\,\,.
\]
\end{itemize}
Setting $V:=\prod_{k=1}^N V_k^*$ will work as desired, since
\begin{eqnarray*}
   \left\| \prod_{k=1}^N \tilde U_k^*-U\right\| &\leq& \sum_{k=N+1}^\infty \| U_k-\idn \| \\
   &\leq& \sum_{k=N+1}^\infty {\rm const}_n\cdot \left(\frac{18}{80}\right)^{\frac k 2} < \frac \eps 2
\end{eqnarray*}
due to Equation~(\ref{EqConstn}) and the proof of Lemma~\ref{LemCompositionUnitary}.
\item Use $V$ to carry out a $\frac\delta 3$-approximation of a unitary extension $\tilde U$ of $U$ on the
state $|\tilde\varphi\rangle$ on the tape (the reason why this is possible
is explained in the proof of Lemma~\ref{LemStandardDecomp}). This results in a vector
$|\varphi\rangle$ with the property that $\|\,|\varphi\rangle-\tilde U|\tilde\varphi\rangle\|<\frac\delta 3$.
\item Simulate $M$ on input $|\varphi\rangle\langle\varphi|$ for $\tau$ time steps
within an accuracy of $\frac\delta 3$, that is, compute an output track state $\rho_{\mathbf O}
\in\mathcal{T}_1^+(\hr_{\mathbf O})$ with $\left\|\rho_{\mathbf O}-M_{\mathbf O}^{\tau}(|\varphi\rangle\langle\varphi|)
\right\|_{\rm Tr}<\frac\delta 3$, move this state to the own output track and halt.
(It has been shown by Bernstein and Vazirani in \cite{BernsteinVazirani} that
there are QTMs that can do a simulation in this way.)
\end{itemize}
Let $\sigma_M:=s_M\otimes \sigma^{(M,n)}$. Using the contractivity of the trace distance with respect to
quantum operations and Lemma~\ref{LemNormInequalities}, we get
\begin{eqnarray*}
   \left\| \mathfrak{U}\left(\sigma_M,\delta\right) \right.&-&\left. M(|\psi\rangle\langle\psi|)\right\|_{\rm Tr}=\\
   &=&\left\|\mathcal{R}(\rho_{\mathbf O})-\mathcal{R}\left(M_{\mathbf O}^\tau(|\psi\rangle
   \langle\psi|)\right)\right\|_{\rm Tr}\\
   &\leq&\left\|\rho_{\mathbf O}-M_{\mathbf{O}}^\tau(|\varphi\rangle\langle\varphi|)\right\|_{\rm Tr}\\
   &&+\left\|M_{\mathbf{O}}^\tau(|\varphi\rangle\langle\varphi|)
   -M_{\mathbf{O}}^\tau(|\psi\rangle\langle\psi|)\right\|_{\rm Tr}\\
&<&\frac\delta 3 +\left\| |\varphi\rangle\langle\varphi|-|\psi\rangle\langle\psi|\right\|_{\rm Tr}\\
   &\leq& \frac\delta 3 +\|\,|\varphi\rangle-|\psi\rangle\|\\
   &\leq& \frac\delta 3 +\|\,|\varphi\rangle-
   \tilde U|\tilde\varphi\rangle\|+\|\tilde U|\tilde\varphi\rangle-|\psi\rangle\|\\
   &<& \frac 2 3 \delta +\left\|\,|\tilde\varphi\rangle-\tilde U^* |\psi\rangle\right\|<\delta\,\,.
\end{eqnarray*}
This proves the claim for pure inputs $\sigma=|\psi\rangle\langle\psi|$. If
$\sigma=\sum_k \lambda_k |\psi_k\rangle\langle \psi_k|$
is a mixed qubit string as explained right before Equation~(\ref{EqExpected}), the result just
proved holds for every convex component of $\sigma$ by the linearity of $M$, i.e.
$\left\| \rho_k-M(|\psi_k\rangle\langle\psi_k|)\right\|_{\rm Tr}<\delta$, and the assertion
of the theorem follows from the joint convexity of the trace distance and the observation
that $\mathfrak U$ takes the same number of time steps for every convex component
$|\psi_k\rangle\langle\psi_k|$.\qed

This proof relies on the existence of a universal QTM $\mathcal U$ in the sense of Bernstein
and Vazirani as given in Equation~(\ref{EqWeakUniversality}). Nevertheless, the proof does not imply
that every QTM that satisfies (\ref{EqWeakUniversality}) is automatically strongly universal
in the sense of Theorem~\ref{MainTheorem1}; for example, we can construct a QTM $\mathcal U$
that always halts after $T$ simulated steps of computation on input $(s_M,T,\delta,|\psi\rangle)$
and that does not halt at all if the input is not of this form. So formally,
\[
   \{{\mathcal U}\mbox{ QTM universal by~(\ref{EqWeakUniversality})}\}
   \supsetneq
   \{{\mathfrak U}\mbox{ QTM strongly universal}\}.
\]

\begin{proposition}[Parameter Strongly Universal QTM]
\label{PropTwoParameter}
\lineclear
There is a fixed-length quantum Turing machine $\mathfrak{U}$ with the property
of Theorem~\ref{MainTheorem1} that additionally satisfies the following:
For every QTM $M$ and
every qubit string $\sigma\in\mathcal{T}_1^+\left(\hr_\s\right)$,
there is a qubit string $\sigma_M\in\mathcal{T}_1^+\left(\hr_\s\right)$
such that
\[
   \left\|\mathfrak{U}\left(\sigma_M,k\right)-M\left(
   \sigma,2 k\right)\right\|_{\rm Tr}<\frac 1 {2k}\qquad
   \mbox{for every }k\in\N
\]
if $M(\sigma,2k)$ is defined for every $k\in\N$, where the length of $\sigma_M$ is bounded
by $\ell(\sigma_M)\leq\ell(\sigma)+c_M$, and $c_M\in\N$ is a constant depending only
on $M$.
\end{proposition}
One might first suspect that this proposition is an easy corollary
of Theorem~\ref{MainTheorem1}, but this is not true. The problem is that
the computation of $M(\sigma,k)$ may take a different number of time steps $\tau$ for different $k$
(typically, $\tau\to\infty$ for $k\to\infty$). Just using the result of Theorem~\ref{MainTheorem1}
would give a corresponding qubit string $\sigma_M$ that depends on $k$, but here we demand
that the qubit string $\sigma_M$ is the {\em same} for every $k$, which is important for
the proof of Theorem~\ref{MainTheorem2} to fit the definition of $QC$.

Thus, we have to give a new proof that is different from the proof of Theorem~\ref{MainTheorem1}.
Nevertheless, the new proof relies essentially on the same ideas and techniques; for this reason,
we will only sketch the proof and omit most of the details.

The proof sketch is based on the idea that a QTM which is universal in the
sense of Bernstein and Vazirani (i.e. as in Equation~(\ref{EqWeakUniversality})) has
a dense set of unitaries that it can apply exactly. We can call such unitaries on $\hr_n$
for $n\in\N$ {\em $\mathfrak U$-exact unitaries}.

This follows from the result by Bernstein and Vazirani that the corresponding UQTM $\mathcal U$
can apply a unitary map $U$ on its tapes within any desired accuracy, if it is
given a description of $U$ as input. It does so by decomposing $U$ into simple (``near-trivial'') unitaries
that it can apply directly (and thus exactly).

We can also call an $n$-block projector $P\in\mathcal{B}(\hr_n)$ $\mathfrak U$-exact if it has
some spectral decomposition $P=\sum_i |\psi_i\rangle\langle\psi_i|$ such that there is a $\mathfrak U$-exact
unitary that maps each $|\psi_i\rangle$ to some computational basis vector of $\hr_n$. If $P$
and $\idn-P$ are $\mathfrak U$-exact projectors on $\hr_n$, then $\mathfrak U$ can do something
like a ``yes-no-measurement'' according to $P$ and $\idn-P$: it can decide whether some vector $|\psi\rangle\in\hr_n$
on its tape is an element of ${\rm ran}\,P$ or of $({\rm ran}\,P)^\perp$ with certainty (if either one
of the two cases is true), just by applying the corresponding $\mathfrak U$-exact unitary, and then
by deciding whether the result is some computational basis vector or another.\\

{\bf Proof Sketch of Proposition~\ref{PropTwoParameter}.}
In analogy to Definition~\ref{DefHaltingQubitStrings}, we can define halting spaces
$\hr_M^{(n)}(t_1,t_2,\ldots,t_j)$ as the linear span of
\begin{eqnarray*}
   H_M^{(n)}(t_1,t_2,\ldots,t_j):=\{|\psi\rangle\in\hr_n\,\,|\,\,
   (|\psi\rangle\langle\psi|,i)\mbox{ is }t_i\mbox{-halting}\\
    \mbox{for }M\,\,(1\leq i\leq j)\}.
\end{eqnarray*}
Again, we have
$\hr_M^{(n)}\left((t_i)_{i=1}^j\right)\perp \hr_M^{(n)}\left((t'_i)_{i=1}^j\right)$ if $t\neq t'$,
and now it also holds that $\hr_M^{(n)}(t_1,\ldots,t_{j},t_{j+1})\subset
\hr_M^{(n)}(t_1,\ldots,t_j)$ for every $j\in\N$.
Moreover, we can define certain $\delta$-approximations $\hr_M^{(n,\delta)}(t_1,\ldots,t_j)$.
We will not get into detail; we will just claim that such a definition can be found in a way
such that these $\delta$-approximations
share enough properties with their counterparts from Definition~\ref{DefApproxHalt}
to make the algorithm given below work.

We are now going to describe how a machine $\mathfrak U$ with the properties given
in the assertion of the proposition works. It expects input of the form
$\left(k,f\otimes s_M\otimes \sigma^{(M,n)}\right)$, where $f\in\{0,1\}$ is a single bit,
$s_M\in\s$ is a self-delimiting description of the QTM $M$, $\sigma^{(M,n)}\in\mathcal{T}_1^+(\hr_\s)$
is a qubit string, and $k\in\N$ an arbitrary integer. For the same reasons as in the proof of
Theorem~\ref{MainTheorem1}, we may without loss of generality assume that the input
is a pure qubit string, so $\sigma^{(M,n)}=|\psi^{(M,n)}\rangle\langle\psi^{(M,n)}|$.
Moreover, due to Lemma~\ref{LemFixedLengthEnough}, we may also assume that $M$ is a fixed-length QTM,
and so $\sigma^{(M,n)}\in\mathcal{T}_1^+(\hr_n)$ is a fixed-length qubit string.

These are the steps that $\mathfrak U$ performs:
\begin{itemize}
\item[(1)] Read the first bit $f$ of the input. If it is a $0$, then proceed with the rest of
the input the same way as the QTM that is given in Theorem~\ref{MainTheorem1}.
If it is a $1$, then proceed with the next step. This ensures that the resulting QTM $\mathfrak U$
will still satisfy the statement of Theorem~\ref{MainTheorem1}.
\item[(2)] Read $s_M$, read $k$, and look for the first blank symbol $\#$ to determine
the length $n:=\ell(\sigma^{(M,n)})$.
\item[(3)] Set $j:=1$ and $\delta_0\in\mathbb{Q}^+$ (depending on $n$) small enough.
\item[(4)] Set $t:=1$.
\item[(5)] Compute $\hr_M^{(n,\delta_0)}(\tau_1,\ldots,\tau_{j-1},t)$. Find a $\mathfrak U$-exact
projector $P_M^{(n)}(\tau_1,\ldots,\tau_{j-1},t)$ with the following properties:
\begin{itemize}
\item[$\bullet$] $P_M^{(n)}(\tau_1,\ldots,\tau_{j-1},t')\cdot P_M^{(n)}(\tau_1,\ldots,\tau_{j-1},t)=0$
for every $1\leq t'<t$,
\item[$\bullet$] $P_M^{(n)}(\tau_1,\ldots,\tau_{j-1},t)\leq P_M^{(n)}(\tau_1,\ldots,\tau_{j-1})$,
\item[$\bullet$] the support of $P_M^{(n)}(\tau_1,\ldots,\tau_{j-1},t)$ is close enough to
$\hr_M^{(n,\delta_0)}(\tau_1,\ldots,\tau_{j-1},t)$.
\end{itemize}
\item[(6)] Make a measurement\footnote{It is not really a measurement, but rather some unitary branching:
if $\psi^{(M,n)}\rangle$ is some superposition in between both subspaces $W:={\rm supp}\left(P_M^{(n)}(\tau_1,\ldots,\tau_{j-1},t)\right)$
and $W^\perp$, then the QTM will do both possible steps in superposition.
} described by $P_M^{(n)}(\tau_1,\ldots,\tau_{j-1},t)$.
If $|\psi^{(M,n)}\rangle$ is an element of the support of $P_M^{(n)}(\tau_1,\ldots,\tau_{j-1},t)$,
then set $\tau_j:=t$ and go to step (7). Otherwise, if $|\psi^{(M,n)}\rangle$ is an element of
the orthogonal complement of the support, set $t:=t+1$ and go back to step (5).
\item[(7)] If $j<2k$, then set $j:=j+1$ and go back to step (4).
\item[(8)] Use a unitary transformation $V$ (similar to the transformation $V$ from the proof
of Theorem~\ref{MainTheorem1}) to do some ``fine-tuning'' on $|\psi^{(M,n)}\rangle$, i.e.
to transform it closer (depending on the parameter $k$) to some space $\tilde\hr_M^{(n)}(\tau_1,\ldots,\tau_j)\supset
\hr_M^{(n)}(\tau_1,\ldots,\tau_j)$ containing the exactly halting vectors.
Call the resulting vector $|\tilde\psi^{(M,n)}\rangle:=V|\psi^{(M,n)}\rangle$.
\item[(9)] Simulate $M$ on input $\left(|\tilde\psi^{(M,n)}\rangle\langle\tilde\psi^{(M,n)}|,2k\right)$ for $\tau_{2k}$
time steps within some accuracy that is good enough, depending on $k$.
\end{itemize}

Let $\tilde\hr_M^{(n,\delta_0)}(t_1,\ldots,t_j)$ be the support of $P_M^{(n)}(t_1,\ldots,t_j)$.
These spaces (which are computed by the algorithm) have the properties
\begin{eqnarray*}
   \tilde\hr_M^{(n,\delta_0)}\left((t_i)_{i=1}^j\right)&\perp& \tilde\hr_M^{(n,\delta_0)}\left((t'_i)_{i=1}^j\right)
   \mbox{ if }t\neq t',\\
   \tilde\hr_M^{(n,\delta_0)}(t_1,\ldots,t_j,t_{j+1})&\subset& \tilde\hr_M^{(n,\delta_0)}(t_1,\ldots,t_j)
   \enspace \forall j\in\N,
\end{eqnarray*}
which are the same as those of the exact halting spaces $\hr_M^{(n)}(t_1,\ldots,t_j)$.
If all the approximations are good enough, then for every $|\psi\rangle\in H_M^{(n)}(t_1,\ldots,t_j)$
there will be a vector $|\psi^{(M,n)}\rangle\in \tilde\hr_M^{(n,\delta_0)}(t_1,\ldots,t_j)$ such
that $\|\,|\psi\rangle-V|\psi^{(M,n)}\rangle\|$ is small. If this $|\psi^{(M,n)}\rangle$ is given
to $\mathfrak U$ as input together with all the additional information explained above, then this algorithm will unambiguously find out
by measurement with respect to the $\mathfrak U$-exact projectors that it computes in step (5) what the
halting time of $|\psi\rangle$ is, and the simulation of $M$ will halt after the correct
number of time steps with probability one and an output which is close to the true output
$M(\sigma,2k)$.\qed\\

{\bf Proof of Theorem~\ref{MainTheorem2}.} First, we use Theorem~\ref{MainTheorem1} to prove
the second part of Theorem~\ref{MainTheorem2}. Let $M$ be an arbitrary QTM,
let $\mathfrak U$ be the (``strongly universal'')
QTM and $c_M$ the corresponding constant from Theorem~\ref{MainTheorem1}.
Let $\ell:=QC_M^\delta(\rho)$, i.e. there
exists a qubit string $\sigma\in\mathcal{T}_1^+(\hr_\s)$ with $\ell(\sigma)=\ell$ such that
\[
   \| M(\sigma)-\rho\|_{\rm Tr}< \delta\,\,.
\]
According to Theorem~\ref{MainTheorem1}, there exists a qubit string $\sigma_M\in\mathcal{T}_1^+(\hr_\s)$
with $\ell(\sigma_M)\leq \ell(\sigma)+c_M=\ell+c_M$ such that
\[
   \|\mathfrak{U}(\sigma_M,\Delta-\delta)-M(\sigma)\|_{\rm Tr}<\Delta-\delta\,\,.
\]
Thus, $\|\mathfrak{U}(\sigma_M,\Delta-\delta)-\rho\|_{\rm Tr}<\Delta$, and
$\ell(\sigma_M,\Delta-\delta)=\ell(\sigma_M)+\ell(\Delta-\delta)\leq \ell+c_M+c_{\delta,\Delta}$,
where $c_{\delta,\Delta}\in\N$ is some constant that only depends on $\delta$ and $\Delta$.
So $QC_{\mathfrak U}^\Delta(\rho)\leq \ell+c_{M,\delta,\Delta}$.

The first part of Theorem~\ref{MainTheorem2} uses Proposition~\ref{PropTwoParameter}. Again, let $M$
be an arbitrary QTM, let $\mathfrak U$ be the strongly universal QTM and $c_M$ the corresponding
constant from Proposition~\ref{PropTwoParameter}. Let $\ell:=QC_M(\rho)$, i.e. there
exists a qubit string $\sigma\in\mathcal{T}_1^+(\hr_\s)$ with $\ell(\sigma)=\ell$ such that
\[
   \| M(\sigma,k)-\rho\|_{\rm Tr}<\frac 1 k\qquad\mbox{for every }k\in\N\,\,.
\]
According to Proposition~\ref{PropTwoParameter}, there exists a qubit string $\sigma_M\in\mathcal{T}_1^+(\hr_\s)$
with $\ell(\sigma_M)\leq\ell(\sigma)+c_M=\ell+c_M$ such that
\[
   \left\|\mathfrak{U}\left(\sigma_M,k\right)-M\left(
   \sigma,2 k\right)\right\|_{\rm Tr}<\frac 1 {2k}\qquad\mbox{for every }k\in\N\,\,.
\]
Thus, $\|\mathfrak{U}(\sigma_M,k)-\rho\|_{\rm Tr}\leq \|\mathfrak{U}(\sigma_M,k)-M(\sigma,2k)\|_{\rm Tr}
+\|M(\sigma,2k)-\rho\|_{\rm Tr}<\frac 1 {2k}+\frac 1 {2k}=\frac 1 k$ for every $k\in\N$. So
$QC_{\mathfrak U}(\rho)\leq \ell+c_M$. \qed

The construction of $\mathfrak U$ is based to a large extent on classical algorithms that
enumerate halting input qubit strings. Since it is in general impossible to decide
unambigously by classical simulation whether some input qubit string $|\psi\rangle$
is perfectly or only approximately halting for a QTM $M$, the UQTM $\mathfrak U$ will also
give some outputs of $M$ which correspond to inputs that are only approximately halting.

With some effort, this observation can be used to generalize the construction of $\mathfrak U$
such that it also captures {\em every} $\eps$-halting input qubit string for $M$ if $\eps>0$ is small
enough, and gives the corresponding output. This leads to the following stability result. A proof
and a more detailed reformulation
can be found in \cite{MeineDiss}.

\begin{theorem}[Halting Stability]
\label{Conjecture}
For every $\delta>0$, there is a computable sequence $a_n(\delta)$ of positive real numbers such that
every qubit string of length $n$ which is $a_n(\delta)$-halting for a QTM $M$ can be enhanced to another
qubit string which is only a constant number of qubits longer, but which makes $\mathfrak U$ halt perfectly and
gives the same output up to trace distance $\delta$.
\end{theorem}

\section{Summary and Perspectives}
\label{SecConclusions}
While Bernstein and Vazirani~\cite{BernsteinVazirani} have defined QTMs with the purpose
to study quantum computational complexity, it has been shown in this paper that QTMs are suitable for
studying quantum algorithmic complexity as well. As proved in Theorem~\ref{MainTheorem1},
there is a universal QTM $\mathfrak U$ that simulates every other QTM until
the other QTM has halted, thereby even obeying the strict halting conditions that
the control is exactly in the halting state at the halting time, and exactly orthogonal
to the halting state before.

Although the calculations in this paper were done for the QTM, it seems plausible
that this construction of a ``strongly universal'' machine can be easily extended to
other models of quantum computation as well. The only assumption is that the quantum computing
device in question computes until it attains some halting state, dependent on the quantum input.

In analogy to the classical situation, this makes it possible to prove that quantum
Kolmogorov complexity depends on the choice of the universal quantum computer only
up to an additive constant, as shown in Theorem~\ref{MainTheorem2}. In the classical
case, this ``invariance property'' turned out to be the cornerstone for the
subsequent development of every aspect of algorithmic information theory. We hope
that the results in this paper will be similarly useful for the development of
a quantum theory of algorithmic information.

There are some more aspects that can be learned from the proofs of Theorems~\ref{MainTheorem1}
and \ref{MainTheorem2}. One example is Lemma~\ref{LemFixedLengthEnough} which essentially
states that indeterminate-length QTMs are no more interesting then fixed-length QTMs, if the length $\ell(\sigma)$
of an input qubit string $\sigma$ is defined as in Definition~\ref{DefQubitStringsandtheirLength}. This supports
the point of view of Rogers and Vedral~\cite{Rogers} to consider the average
length $\bar\ell(\sigma)$ instead, that is, the expectation value of the length $\ell$.
If the halting of the underlying quantum computer is still defined as in this paper,
then our result applies to their definition, too.

The construction of the strongly universal QTM $\mathfrak U$ in the proof of Theorem~\ref{MainTheorem1}
is such that $\mathfrak U$ starts with a completely classical computation, followed by the
application of classically selected unitary operations. But the same steps (on the same input) can be done
by a machine that has a purely classical control, selecting at each step of the computation
a unitary transformation that is applied to an unknown quantum state (that was part of the input)
without any measurement. Thus, it seems that at least from the point of view of quantum Kolmogorov complexity
$QC^\delta$,
it is sufficient to consider machines with a completely classical control. Such machines do not
have the problem of ``approximate halting'' described in Subsection~\ref{SubsecIntroQTM}.

There may be interesting applications of extending algorithmic information theory to the quantum case.
One exciting perspective is that in a quantum theory of algorithmic complexity,
both the inherent notions of ``randomness'' of quantum theory and ``algorithmic randomness''
originating from undecidability results will occur (and maybe be related) in a single theory.
One possible application of quantum Kolmogorov complexity might be
to analyze a fully quantum version of the thought experiment of Maxwell's demon in
statistical mechanics, since its classical counterpart has already proved useful for the corresponding
classical analysis (cf. \cite{Vitanyibook}).

\appendix
\begin{lemma}[Inner Product and Dimension Bound]
\label{LemInnerProduct}
\lineclear
Let $\hr$ be a Hilbert space, and let $|\psi_1\rangle,\ldots,|\psi_N\rangle\in\hr$ with
$\|\,|\psi_i\rangle\|=1$ for every $i\in\{1,\ldots,N\}$, where $2\leq N\in\N$. Suppose that
\[
   \left\vert\strut\langle \psi_i|\psi_j\rangle\right\vert<\frac 1 {N-1}\qquad\mbox{for every }i\neq j\,\,.
\]
Then, $\displaystyle \dim\hr\geq N$.
\end{lemma}

{\bf Proof.} We prove the statement by induction in $N\in\N$. For $N=2$, the statement
of the theorem is trivial. Suppose the claim holds for some $N\geq 2$, then consider
$N+1$ normalized vectors $|\psi_1\rangle,\ldots,|\psi_{N+1}\rangle\in\hr$, where $\hr$ is an arbitrary
Hilbert space.
Suppose that $|\langle \psi_i|\psi_j\rangle|<\frac 1 N$ for every $i\neq j$.
Let $P:=\idn-|\psi_{N+1}\rangle\langle\psi_{N+1}|$, then $P|\psi_i\rangle\neq 0$ for every
$i\in\{1,\ldots,N\}$, and let
\[
   |\varphi_i'\rangle:=P |\psi_i\rangle\,\,,\qquad
   |\varphi_i\rangle:=\frac{|\varphi_i'\rangle}{\|\,|\varphi_i'\rangle\|}\,\,.
\]
The $|\varphi_i\rangle$ are normalized vectors in the Hilbert
subspace $\tilde\hr:={\rm ran}(P)$ of $\hr$. Since $\|\,|\varphi_i'\rangle\|^2
=\langle \psi_i|\psi_i\rangle-|\langle\psi_i|\psi_{N+1}\rangle|^2
   >1-\frac 1 {N^2}$,
it follows that the vectors $|\varphi_i\rangle$ have small inner product: Let $i\neq j$, then
\begin{eqnarray*}
   |\langle \varphi_i | \varphi_j\rangle|&=&\frac 1 {\|\,|\varphi_i'\rangle\|\cdot\|\,|\varphi_j'\rangle\|}
   |\langle \varphi_i'|\varphi_j'\rangle|\\
   &<& \frac {\left\vert
      \langle \psi_i|\psi_j\rangle- \langle\psi_{N+1}|\psi_j\rangle\langle\psi_i|\psi_{N+1}\rangle
   \right\vert}
   {\sqrt{1-\frac 1 {N^2}}\sqrt{1-\frac
   1 {N^2}}}\\
   &<& \frac 1 {1-\frac 1 {N^2}} \left(\frac 1 N + \frac 1 {N^2}\right) = \frac 1 {N-1}\,\,.
\end{eqnarray*}
Thus, $\dim\tilde\hr\geq N$, and so $\dim\hr\geq N+1$.\qed

\begin{lemma}[Composition of Unitary Operations]
\label{LemCompositionUnitary}
\lineclear
Let $\hr$ be a finite-dimensional Hilbert space, let $(V_i)_{i\in\N}$ be a sequence
of linear subspaces of $\hr$ (which have all the same dimension), and let $U_i:V_i\to V_{i+1}$
be a sequence
of unitary operators on $\hr$ such that $\sum_{k=1}^\infty \| U_k-\idn\|$ exists.
Then, the product $\prod_{k=1}^\infty U_k=\ldots\cdot U_3\cdot U_2\cdot U_1$ converges
in operator-norm to an isometry $U: V_1\to\hr$.
\end{lemma}
{\bf Proof.} We first show by induction that $\left\| \prod_{k=1}^N U_k - \idn\right\|
\leq \sum_{k=1}^N \| U_k-\idn\|$. This is trivially true for $N=1$; suppose it is true
for $N$ factors, then
\begin{eqnarray*}
   \left\| \prod_{k=1}^{N+1} U_k - \idn\right\|&\leq&\left\|
    \prod_{k=1}^{N+1} U_k-\prod_{k=1}^N U_k\right\| +
    \left\| \prod_{k=1}^N U_k-\idn\right\|\\
    &\leq& \left\|(U_{N+1}-\idn)\prod_{k=1}^N U_k\right\|
    +\sum_{k=1}^N \| U_k-\idn\|\\
    &\leq&\sum_{k=1}^{N+1} \| U_k-\idn\|\,\,.
\end{eqnarray*}

By assumption, the sequence $a_n:=\sum_{k=1}^n\| U_k-\idn\|$ is a Cauchy sequence; hence, for every
$\eps>0$ there is an $N_\eps\in\N$ such that for every $L,N\geq N_\eps$ it holds that
$\sum_{k=L+1}^N \| U_k-\idn\|<\eps$. Consider now the sequence $V_n:=\prod_{k=1}^n U_k$.
If $N\geq L\geq N_\eps$, then
\begin{eqnarray*}
   \| V_N-V_L\| &=& \left\| \prod_{k=L+1}^N U_k \cdot \prod_{k=1}^L U_k - \prod_{k=1}^L U_k\right\|\\
   &\leq&\left\| \prod_{k=L+1}^N U_k - \idn\right\| \cdot \left\| \prod_{k=1}^L U_k\right\|\\
   &\leq& \sum_{k=L+1}^N \| U_k-\idn\| <\eps\,\,,
\end{eqnarray*}
so $(V_n)_{n\in\N}$ is also a Cauchy sequence and converges in operator norm to
some linear operator $U$ on $V_1$. It is easily checked that $U$ must be isometric.\qed

\begin{lemma}[Norm Inequalities]
\label{LemNormInequalities}
Let $\hr$ be a finite-dimensional Hilbert space, and
let $|\psi\rangle,|\varphi\rangle\in\hr$ with $\|\,|\psi\rangle\|=\|\,|\varphi\rangle\|=1$.
Then,
\[
   \|\,|\psi\rangle\langle\psi|-|\varphi\rangle\langle\varphi|\,\|_{\rm{Tr}}
   \leq \|\,|\psi\rangle-|\varphi\rangle\|\,\,.
\]
Moreover, if $\rho,\sigma\in\mathcal{T}_1^+(\hr)$ are density operators, then
\[
   \|\rho-\sigma\|\leq \|\rho-\sigma\|_{\rm Tr}\,\,.
\]
\end{lemma}
{\bf Proof.} Let $\Delta:=|\psi\rangle\langle\psi|-|\varphi\rangle\langle\varphi|$.
Using \cite[9.99]{NielsenChuang},
\begin{eqnarray*}
   \|\Delta\|_{\rm Tr}^2&=&1-|\langle\psi|\varphi\rangle|^2
   =\left(\strut 1-|\langle\psi|\varphi\rangle|\right)
   \underbrace{\left(\strut 1+|\langle\psi|\varphi\rangle|\right)}_{\leq 2}\\
   &\leq&2-2|\langle\psi|\varphi\rangle|\leq 2-2{\rm Re}\langle\psi|\varphi\rangle\\
   &=&\langle\psi-\varphi|\psi-\varphi\rangle=\|\,|\psi\rangle-|\varphi\rangle\|^2\,\,.
\end{eqnarray*}

Let now $\tilde\Delta:=\rho-\sigma$, then
$\tilde\Delta$ is Hermitian. We may assume that one of its eigenvalues which has largest absolut
value is positive (otherwise interchange $\rho$ and $\sigma$), thus
\begin{eqnarray*}
   \|\tilde\Delta\|&=&
   \max_{\|\,|v\rangle\|=1}\langle v|\tilde\Delta |v\rangle
   =\max_{P\mbox{ proj., } {\rm Tr } P=1}{\rm Tr}(P\tilde\Delta)\\
   &\leq& \max_{P\mbox{ proj.}} {\rm Tr}(P\tilde\Delta)=\|\tilde\Delta\|_{\rm Tr}
\end{eqnarray*}
according to \cite[9.22]{NielsenChuang}.
\qed

\begin{lemma}[Dimension Bound for Similar Subspaces]
\label{LemDimBoundSimilar}
\lineclear
Let $\hr$ be a finite-dimensional Hilbert space, and let $V,W\subset\hr$
be subspaces such that for every $|v\rangle\in V$ with $\|\,|v\rangle\|=1$
there is a vector $|w\rangle\in W$ with $\|\,|w\rangle\|=1$ which
satisfies $\|\,|v\rangle-|w\rangle\|\leq\eps$, where
$0<\eps\leq \frac 1 {4(\dim V -1)^2}$ is fixed.
Then, $\dim W\geq \dim V$. Moreover, if additionally $\eps\leq \frac 1 {36} \left(\frac 5 2\right)
^{2-2\dim V}$ holds, then
there exists an isometry $U:V\to W$ such that
$\|U-\idn\|<\frac 8 3 \sqrt{\eps}\left(\frac 5 2\right)^{\dim V}$.
\end{lemma}
{\bf Proof.} Let $\{|v_1\rangle,\ldots,|v_d\rangle\}$ be an orthonormal basis of $V$.
By assumption, there are normalized vectors $\{|w_1\rangle,\ldots,|w_d\rangle\}\subset W$ with
$\|\,|v_i\rangle-|w_i\rangle\|\leq\eps$ for every $i$.
From the definition of the trace distance for pure states (see \cite[(9.99)]{NielsenChuang}
together with Lemma~\ref{LemNormInequalities},
it follows for every $i\neq j$
\begin{eqnarray*}
   \sqrt{1-|\langle w_i|w_j\rangle|^2}&=& \|\,|w_i\rangle\langle w_i|-|w_j\rangle\langle w_j|\,\|_{\rm Tr}\\
   &\geq& \|\,|v_i\rangle\langle v_i|-|v_j\rangle\langle v_j|\,\|_{\rm Tr}\\
   && -\|\,|v_i\rangle\langle v_i|-|w_i\rangle\langle w_i|\,\|_{\rm Tr}\\
   && -\|\,|v_j\rangle\langle v_j|-|w_j\rangle\langle w_j|\,\|_{\rm Tr}\\
   &\geq&1-\|\,|v_i\rangle-|w_i\rangle\|-\|\,|v_j\rangle-|w_j\rangle\|\\
   &\geq& 1-2\eps\,\,.
\end{eqnarray*}
Thus, $|\langle w_i|w_j\rangle| < 2 \sqrt\eps\leq \frac 1 {d-1}$, and it follows
from Lemma~\ref{LemInnerProduct} that $\dim W\geq d$. Now apply the Gram-Schmidt
orthonormalization procedure to the vectors $\{|w_i\rangle\}_{i=1}^d$:
\begin{eqnarray*}
   |\tilde e_k\rangle &:=& |w_k\rangle - \sum_{i=1}^{k-1} \langle w_k|e_i\rangle |e_i\rangle\,\,,\qquad
   |e_k\rangle := \frac{|\tilde e_k\rangle}{\|\,|\tilde e_k\rangle\|}\,\,.
\end{eqnarray*}
Use $\left|\strut \|\,|\tilde e_k\rangle\|-1\right|=\left|\strut \|\,|\tilde e_k\rangle\|
-\|\,|w_k\rangle\|\right|\leq \|\,|\tilde e_k\rangle-|w_k\rangle\|$ and calculate
\begin{eqnarray*}
   \|\,|\tilde e_k\rangle-|w_k\rangle\|&=&\left\| \sum_{i=1}^{k-1} \frac{\langle w_k|\tilde e_i\rangle
   |\tilde e_i\rangle}{\|\,|\tilde e_i\rangle\|^2}\right\|\\
   &\leq& \sum_{i=1}^{k-1}\frac{\left| \langle w_k|\tilde e_i-w_i\rangle\right| + \left|
   \langle w_k|w_i\rangle\right|}{\|\,|\tilde e_i\rangle\|}\\
   &\leq& \sum_{i=1}^{k-1} \frac{\|\,|\tilde e_i\rangle-|w_i\rangle\|+2\sqrt{\eps}}
   {1-\|\,|\tilde e_i\rangle-|w_i\rangle\|}\,\,.
\end{eqnarray*}
Let $\Delta_k:=\|\,|\tilde e_k\rangle-|w_k\rangle\|$ for every $1\leq k \leq d$. We will now
show by induction that $\Delta_k\leq 2 \sqrt{\eps}\left[ \frac 2 5\left(\frac 5 2\right)^k-1\right]$.
This is trivially true for $k=1$, since $\Delta_1=0$. Suppose it is true for every $1\leq i \leq k-1$,
then in particular, $\Delta_i\leq \frac 1 3$ by the assumptions on $\eps$ given in the statement of
this lemma, and
\begin{eqnarray*}
   \Delta_k&\leq& \sum_{i=1}^{k-1}\frac{\Delta_i+2\sqrt{\eps}}{1-\Delta_i}\\
   &\leq&  \frac 3 2 \sum_{i=1}^{k-1} \left(
      2\sqrt{\eps}\left[\frac 2 5\left(\frac 5 2 \right)^i -1\right]+2\sqrt{\eps}
   \right)\\
   &=&2 \sqrt{\eps}\left[\frac 2 5 \left(\frac 5 2\right)^k -1\right]\,\,.
\end{eqnarray*}
Thus, it holds that
\begin{eqnarray*}
   \|\,|e_k\rangle-|v_k\rangle\|&\leq&\|\,|e_k\rangle-|\tilde e_k\rangle\|\\
   &&+\|\,|\tilde e_k\rangle-|w_k\rangle\|+\|\,|w_k\rangle-|v_k\rangle\|\\
   &\leq& 2\|\,|\tilde e_k\rangle-|w_k\rangle\|+\eps\\
   &\leq& 4\sqrt{\eps}\left[\frac 2 5 \left(\frac 5 2\right)^k -1\right]+\eps\,\,.
\end{eqnarray*}
Now define the linear operator $U:V\to W$ via linear extension of
$U|v_i\rangle:=|e_i\rangle$ for $1\leq i\leq d$. This map is an isometry, since it
maps an orthonormal basis onto an orthonormal basis of same dimension.
By substituting $|v\rangle=\sum_{k=1}^d \alpha_k |v_k\rangle$ and using $\eps<4\sqrt{\eps}$
and the geometric series, it easily follows that $\|\,U|v\rangle-|v\rangle\|\leq \frac 8 3
\sqrt{\eps}\left(\frac 5 2 \right)^d$ if $\|\,|v\rangle\|=1$.\qed

\begin{lemma}[Stability of the Control State]
\label{LemStability}
\lineclear
If $|\psi\rangle,|\varphi\rangle,|v\rangle\in\cn$ and $\|\,|\psi\rangle\|=\|\,|\varphi\rangle\|=1$
and $|v\rangle\neq 0$, then
it holds for every QTM $M$ and every $t\in\N_0$
\begin{eqnarray*}
   &&\left| \langle q_f | M_{\mathbf{C}}^t(|\psi\rangle\langle\psi|)|q_f\rangle-
   \langle q_f | M_{\mathbf{C}}^t(|\varphi\rangle\langle\varphi|)|q_f\rangle\right|\\
   &&\qquad\leq\left\|\strut\, |\psi\rangle\langle\psi|-|\varphi\rangle\langle\varphi|\,\right\|_{\rm Tr}\,\,,\\
   &&\left|\strut \langle q_f|M_{\mathbf{C}}^{t}(|v\rangle\langle v|)|q_f\rangle-
   \langle q_f|M_{\mathbf{C}}^{t}(|v^0\rangle\langle v^0|)|q_f\rangle\right|\\
   &&\qquad\leq
   \left|\strut 1-\|\,|v\rangle\|^2\right|
   \,\,.
\end{eqnarray*}
\end{lemma}
{\bf Proof.} Using the Cauchy-Schwarz inequality, Lemma~\ref{LemNormInequalities} and
the contractivity of quantum operations
with respect to the trace distance (cf. \cite[(9.35)]{NielsenChuang}), we get
the chain of inequalities
\begin{eqnarray*}
   \Delta_t&:=&
   \left| \langle q_f | M_{\mathbf{C}}^t(|\psi\rangle\langle\psi|)|q_f\rangle-
   \langle q_f | M_{\mathbf{C}}^t(|\varphi\rangle\langle\varphi|)|q_f\rangle\right|\\
   &\leq& \left\|\,M_{\mathbf{C}}^t\left(\strut|\psi\rangle\langle\psi|\right)
   -M_{\mathbf{C}}^t\left(\strut|\varphi\rangle\langle\varphi|\right)
   \right\|\\
   &\leq& \left\|\,M_{\mathbf{C}}^t\left(\strut|\psi\rangle\langle\psi|\right)
   -M_{\mathbf{C}}^t\left(\strut|\varphi\rangle\langle\varphi|\right)
   \right\|_{\rm Tr}\\
   &\leq&\left\|\strut |\psi\rangle\langle\psi|-|\varphi\rangle\langle\varphi|\right\|_{\rm Tr}\,\,.
\end{eqnarray*}
The second inequality can be proved by an analogous calculation.
\qed

\section*{Acknowledgment}
Sincere thanks go to Wim van Dam, Caroline Rogers, Torsten Franz, and David Gross for helpful discussions,
and to an anonymous referee for very useful comments on a previous draft.
Also, the author would like to thank his collegues Ruedi Seiler, Arleta Szko\l a,
Rainer Siegmund-Schultze, Tyll Kr\"uger, and Fabio Benatti for constant support and encouragement.

\end{document}